\documentclass[12pt]{article}
\usepackage{amsmath}
\usepackage{graphicx}
\usepackage{natbib}
\usepackage{url} % not crucial - just used below for the URL 

\newcommand{\blind}{0}

\usepackage{multicol, multirow}
\usepackage{subcaption}

\graphicspath{{Figures_sel/}}
\newlength{\figurewidth} 
\newlength{\figureheight} 

\begin{document}

\def\spacingset#1{\renewcommand{\baselinestretch}%
{#1}\small\normalsize} \spacingset{1}

%%%%%%%%%%%%%%%%%%%%%%%%%%%%%%%%%%%%%%%%%%%%%%%%%%%%%%%%%%%%%%%%%%%%%%%%%%%%%%

\if0\blind
{
  \title{\bf A Dynamic Bayesian Model for Interpretable Decompositions of Market Behaviour}
  \author{Th\'{e}ophile Griveau-Billion\\
    Department of Statistics, University of Imperial College London\\
    and \\
    Ben Calderhead\\
    Department of Statistics, University of Imperial College London}
  \maketitle
} \fi

\if1\blind
{
  \bigskip
  \bigskip
  \bigskip
  \begin{center}
    {\LARGE\bf A Dynamic Bayesian Model for Interpretable Decompositions of Market Behaviour}
\end{center}
  \medskip
} \fi

\bigskip
\begin{abstract}
% Should be less than 200 words
We propose a heterogeneous simultaneous graphical dynamic linear model (H-SGDLM), which extends the standard SGDLM framework to incorporate a heterogeneous autoregressive realised volatility (HAR-RV) model.  This novel approach creates a GPU-scalable multivariate volatility estimator, which decomposes multiple time series into economically-meaningful variables to explain the endogenous and exogenous factors driving the underlying variability.  This unique decomposition goes beyond the classic one step ahead prediction; indeed, we investigate inferences up to one month into the future using stocks, FX futures and ETF futures, demonstrating its superior performance according to accuracy of large moves, longer-term prediction and consistency over time.
\end{abstract}

\noindent%
{\it Keywords:}  Dynamic Bayesian model, dynamic graphical model, GPU computation, market-stress forecasting, sparse multivariate model, volatility forecasting.
\vfill

{\it Word count:} $11912$

\newpage
\spacingset{1.5} % DON'T change the spacing!

%%%%%%%%%%%%%%%%%%%%%%%%%%%%%%%%%%%%%%%%%%%%%%%%%%%%%%%%%%%%%%%%%%%%%%%%%%%%%%%%%%%%
\section{Introduction}
\label{sec:intro}
The behaviour of each asset in the market is driven by both endogenous factors representing the information specific to that asset and exogenous factors representing the impact of the market.
The heterogeneous market hypothesis considers that agents in the market trade with different objectives. While these objectives can be related to many characteristics, \cite{Muller:1997he} argue that most of them are reflected in the time horizon and highlight this fact by studying the impact of heterogeneous investment horizons with a volatility model using the returns computed at different frequencies.  Following this reasoning, the study of a time series at different frequencies should reflect the impact of endogenous factors on that asset.
On the other hand, for each asset the exogenous variables correspond to the time series that have the greatest impact on the behaviour of that asset. The classic approach involves creating a graph of cross-series relationships between the time series in a market based on the covariance matrix. However, the covariance is a symmetric object, while in reality some assets might influence many and others none, in a non-symmetric manner. Hence, a model that selects the exogenous factors without assuming such symmetry in the relationship may be more appropriate.
Combining these two sources of driving factors, our aim is to construct a model that decomposes each time series into its endogenous and exogenous parts. Having such a decomposition gives us a better understanding of what is driving each time series’ behaviour and thus the market as a whole. A better structural understanding both for a single asset and the overall market allows us to produce more accurate inferences and stress indicators.\bigskip

%\subsection*{The model we propose in this report}
We propose a model that extends the Simultaneous Graphical Dynamic Linear Model (SGDLM) of \cite{Gruber:2016wn, Gruber:2016il, Zhao:2016kr, McAlinn:2016tm} to incorporate the heterogeneous autoregressive realised volatility (HAR-RV) model of \cite{Corsi:2004eg, Corsi:2008ei, Corsi:2009ue}. Combining the HAR-RV model with the SGDLM framework creates an easy-to-scale multivariate volatility estimator. Each time series of daily log-volatility is a DLM with idiosyncratic factors from the HAR-RV model, cross-series relationship factors from the multivariate Wishart and any additional variables specific to this time series. The variables can be clustered into two groups: endogenous and exogenous. The endogenous group represents the information specific to the stock, while the exogenous one represents the influence of the environment. As a result of the flexibility of the SGDLM framework, as long as the normality condition of each DLM is respected they can easily be extended to include additional variables, and these do not have to be the same for every series. Hence, we will present different extensions that move further away from the standard HAR-RV formulation .\bigskip

The decomposition performed by our proposed model can explain at any time which economic variables are likely to be driving the variability; for example, it may be due to the sector, the market or internal information, or at a particular frequency. This decomposition of the move into economically-meaningful variables and the capacity of the algorithm to follow their evolution creates new signals to study. In order to get the most out of these different signals we used a simple scale space change point algorithm, such that the correlations between these signals and the underlying time series allows us to perform more reliable inference for days and weeks ahead for each stock, and indeed the whole market. In addition, this model has proven to be an efficient market stress indicator and forecaster.  While \cite{Gruber:2016wn} observed striking similarities between some metric of the SGDLM and the market stress index St. Louis Fed Financial Stress Index, our model appears to offer insight into the likely moves of this index weeks ahead.\bigskip

%\subsection*{An overview of results using our novel approach}
In order to assess the performance of the HSGDLM model we look at the percentage of measured points that lie within the inferred confidence interval. We are especially interested by the figures obtained for large moves in the variance, in particular positive moves, since more than $68\%$ of them are negative. We compare the percentage of correct predictions between algorithms for a confidence interval smaller than the move. If our novel approach for decomposing the variance into different groups of variables representing complementary information at different scales works, it should give us new insights on what is driving the market and thus be much better at predicting large moves. Indeed, when considering the realized variance of stocks in a group of $487$ European stocks over $18$ years from $2001$ to $2019$, the HAR-RV and SGDLM models correctly predicted only respectively $53.24\%$ and $34.69\%$ of the changes in variance bigger than $9.28\%$. In contrast, our model inferred $63.89\%$ of them correctly and in a different environment with stocks from the S\&P$500$ over the same period, with the same model and parameters, that figure is $65.93\%$.\bigskip

In addition, the benefits of these added factors are not related to any specific market situation, since we show with a backtest that the performance of the new signals is constant through time. Instead of relying only on the direct one-step ahead inference from the multivariate model we leverage the fact that the HSGDLM model sequentially updates the coefficients with the Kalman Filter equation to study the evolution of these coefficients with time. More precisely, we use the multivariate HSGLDM model as a feature extractor to create time series of these coefficients. Then, we group the coefficients depending on the economic information they represent, e.g. endogenous or exogenous, and study this newly created time series. By studying the evolution of the time series of coefficients with a change point algorithm we can go beyond the one step ahead prediction and perform much longer term predictions such as one month ahead. While the performance of long term prediction decreases compared to shorter term ones they observe the same consistency through time.\bigskip

In addition, since each group of coefficient represents a specific economic factor, our model can answer more interesting structural questions such as: does the variance increase when the spread between the exogenous and endogenous variables increases or decreases? Furthermore, the flexibility of the statistical tools used by the SGDLM framework and the economic decomposition into endogenous and exogenous groups allows our approach to be applied to other financial metrics. We will demonstrate applications on modelling the prices of the ETF and FX futures with the same performance qualities: accuracy of large moves, longer-term prediction and consistency through time.\bigskip

The literature review is in Section \ref{sec:lit_rev}. Section \ref{sec:variogram} describes the variogram study that provided the underlying motivation for our research. The SGDLM model is described in section \ref{sec:SGDLM} and the HAR-RV model in section \ref{sec:HAR}. Section \ref{sec:prop} explains the combined model and a variety of implementation issues. Section \ref{sec:results} shows the results on stocks from the S\&P$500$ and Section \ref{sec:resultsDiffEnv} on other environments and assets.

\section{Literature review}
\label{sec:lit_rev}
\subsection{The limitations of GARCH models}
\cite{Muller:1997he} proposed the HARCH variance model to build on the idea of heterogeneous investment horizons. Following their approach, we start by working with variances before extending the model to other metrics. The variance of a time series of asset prices has many interesting properties that many models have previously tried to capture. Some stylized facts of variance-processes are of particular interest to us. The first is the time asymmetry observed in financial time series, i.e. the importance of distinguishing past from future. From this we can conclude that a model of variance should not be time reversible. This assumption then leads to a leverage effect, which is the response asymmetry between the magnitude of previous stock returns and future variance.  Furthermore, we may consider multifractality, which defines how the distribution of returns should change when studied at different time scales. These stylized facts are not reproduced by the widely used stochastic volatility or GARCH models. Further complicating matters, the distribution of each returns time series may have different characteristics and these can evolve through time. Papers such as \cite{Muller:1997he}, \cite{Muzy:2000cz}, \cite{Lynch:2003cn} studied the multifractal behaviour and time asymmetry of financial returns. In \cite{Calvet:2001fo, Calvet:2004fm} the authors extended the multifractal model of asset returns introduced by Mandelbrot to propose a stationary volatility model that can be estimated with maximum likelihood. Their Markov switching volatility model is based on a multiplicative cascade of volatility components, each representing a different frequency. While this model has a closed form likelihood it still requires an optimization step and the parameters lack clear interpretation. In contrast, we seek a model that helps us understand the time series, such that all parameters and variables have a clear interpretation.

\subsection{The link between physics and volatility}
As explained by the heterogeneous market hypothesis in \cite{Muller:1997he}, different market participants trade with different objectives. One of them is the time scale and it impacts participants asymmetrically; the proposed HARCH model is a modified GARCH model using returns at different frequencies. They used this decomposition to show that lower time scales influence higher scales more than the other way around. In other words, short term traders are more greatly impacted by trades from long term traders than the other way around. This information asymmetry between scales has been further detailed in many papers by studying the volatility cascade, i.e. the flow of information between scales, see \cite{Zumbach:2001iz} for a study of this cascade and its parallel with ideas from physics. Indeed, the name came from a similar concept in physics where the fluid vorticity cascade observed in turbulent states has a comparable behaviour to the distribution of financial returns at different scales, thus allowing the same mathematics to be used for volatility modelling. See also for example \cite{Muzy:2000cz} for a multifractal random walk model that produces a scale-invariant stochastic volatility model. This form of statistical feedback model led \cite{Borland:1998wf, Borland:2010km, Borland:2004wu, Borland:2005vd, Borland:2005tk} to propose a process with a noise component following the non-linear Fokker-Planck equation, the solution of which results in a q-Gaussian or Tsallis distribution. With this approach the model could theoretically give a distribution that models the behaviour of the price process for different frequencies and thus characterize each stock into different categories depending on the value of $q$. However, although this complicated distribution has interpretable parameters, it is hard to fit with few data points.

\subsection{The HAR-RV model of Corsi}
The models described previously are able to reproduce most stylized facts, especially multifractality, however they are hard to estimate and can lack clear economic interpretation. While many papers make use of multiplicative models, such as \cite{Calvet:2004fm}, for reproducing multifractality, \cite{LeBaron:2001cp} proved for the first time that a simple three-factor additive model could also display this behaviour. Based on these results, \cite{Corsi:2004eg, Corsi:2008ei, Corsi:2009ue} build the HAR-RV model which uses the heterogeneous market idea to create a simple additive volatility model combining $AR(1)$ processes at different scales. This model is able to reproduce multifractal scaling while conserving simple economic interpretations and being easily extendible to incorporate additional stylized facts such as leverage and jumps \cite{Corsi:2009ue}. They used this decomposition into different time-scales to learn which economic factor is influencing the time series. But this was done for each stock individually and lacks multivariate connections.

\subsection{How can we model cross-series relationships?}
In order to obtain a model of the market as a whole it is important to take into account the cross-series relationships between volatilities. Many studies have shown the clear performance increase of integrating the multivariate effect in their model, such as multivariate- GARCH and -HEAVY models, see e.g. \cite{Noureldin:2011bc}, however such models use heavy MCMC parametrization techniques which make them expensive to scale. Other papers proposed methods to minimize this cost; for example \cite{Nakajima:2014dt} proposed a multivariate SV model using a covariance matrix Cholesky decomposition to allow for parallel evaluation of independent univariate SV models but still fit the model with an expensive MCMC procedure. Another limitation of most multivariate models is their dependence on a historical window to perform the parametrization compared to employing a sequential learning approach. One reason for this is the complexity of online learning algorithms for non-Gaussian models, see e.g. \cite{mj:2006vf}, which presents an auxiliary particle filter to sequentially update a stochastic volatility model with jumps. In contrast, our proposed model is multivariate and due to a Gaussian setup it can sequentially adapt its parameters sequentially to the evolving market.

\subsection{The SGDLM model}
The novel model we present here aims at explaining what drives the time series. This approach will give both new structural information about the market, as well as better longer-term forecasts. We build each module of the model to capture the different observations described above, and then combine each time series into a multivariate model of the market under study. This is possible using the SGDLM framework developed by \cite{Gruber:2016wn, Gruber:2016il, Zhao:2016kr, McAlinn:2016tm}, which considers an environment of independent time series modelled by a simple Normal DLM, using a sparse covariance matrix representation to incorporate cross-series relationships. Also, the SGDLM does not follow the classic multivariate variance approach of assuming symmetric connections between assets by using a covariance matrix to model the cross-series relationship. Instead, the SGDLM has a multivariate Wishart model running in parallel to sequentially select for each time series which other assets have influence on it. After this selection phase, the candidates are tested as a regression variable. This techniques creates non-symmetric cross-series relationships which produce interesting information on the underlying market risks.\bigskip

In addition, compared to the previously described multivariate approaches, the combination of de-coupling and re-coupling steps allows for parallel updates of each DLM with Kalman filter equations. This highly parallel architecture to sequentially update the series in a large scale environment is perfectly suited for a GPU. Since the DLMs are independent they do not need to follow the same model; their only requirement is for the distributions of the states and observations to be Normally distributed. Since this sequential approach continuously updates the coefficients of the DLMs, their distribution will adapt to changes in the environment.

%%%%%%%%%%%%%%%%%%%%%%%%%%%%%%%%%%%%%%%%%%%%%%%%%%%%%%%%%%%%%%%%%%%%%%%%%%%%%%%%%%%%
\section{The variogram study}
\label{sec:variogram}
\subsection{Some insight from variograms}
The motivation for this paper started with a simple observation. In an article written for Risk, \cite{Anonymous:neN5kQNv} used a volatility signature plot to determine the appropriate frequency to use for computing the realized variance (RV) over high frequency returns. This plot corresponded to the realized variance averaged at different time scales, which they then used to determine the time scale on which the microstructure effects stop interfering on the realized variance. On this graphic they also observed distinct patterns for liquid and illiquid stocks. In the first case the realized variance was increasing with the scale while it was decreasing for the illiquid stocks. A similar approach was used in \cite{Borland:2004wu} and \cite{Borland:2005tk} where they used the variogram computed with a stock’s returns to exhibit the properties of multifractality and correlation of volatilities across time scales. Their variogram corresponded to the averaged squared difference between squared returns at different scales. Both of these approaches used high frequency data, but what would we see on lower frequency scales, such as daily returns? Following this logic we computed the averaged realized variance for different frequencies of returns ranging from days to weeks, rescaled by frequency. The common stochastic diffusion process of log-prices assumes a property of self-similarity. If that was correct, as explained in \cite{Borland:2004wu}, the variance should be scale-invariant and thus exhibit a straight line on the variogram. In practice we observe three different clusters of stocks on this plot. Stocks with variances that increase, stay constant or decrease w.r.t. the time scale, i.e. going from daily to weekly returns. Can we explain this behaviour? As detailed in the following section, we can indeed formulate different economic interpretations which lead us to follow this logic further.

\subsection{An economic phenomenon we want to capture}
We expect to detect with a variogram what practitioners often describe as the gamma effect. Gamma corresponds here to the Greek letter $\Gamma$ which represents for option hedgers the derivative of the delta. The delta $\Delta$ corresponds to the first derivative of the price of the option with respect to the price of the underlying asset and aims to measure the sensitivity of the option to the price of the asset. Hence the gamma represents the second order sensitivity to variations in the price of the asset.\bigskip

Let us consider stocks; traders have to delta-hedge their positions, which signifies buying or selling a certain quantity of stocks in order for their $\Delta$ to be zero, and to do so they follow the results given by the equations of Black-Scholes. However, as they are not alone in the market a stock’s price might not move in their direction. When $\Gamma$ is positive, if the price increases then the $\Delta$ will increase too; while for a negative $\Gamma$ the $\Delta$ will decrease. Thus, depending on the $\Gamma$ traders will not react symmetrically to the move of the stock. Furthermore, when a trader wants to delta-hedge his position this second order effect might have an unexpected impact if the absolute value of $\Gamma$ is high, since it means a high sensitivity on the variation of the stocks. This is a simple description, for more details on Delta and Gamma hedging refer to \cite{Hull:2019wk}.\bigskip

If we consider a stock on which a trader has a big option position its hedging might have a detectable impact on the time series of prices. Indeed, in the case where the trader tries to keep the price in a specific range, we might not see any impact on the daily scale, but on a lower one, such as weekly returns, we might observe a variance lower than usual due to this bounding effect.

\subsection{Description of the approach}
The proof of concept for our intuition uses the following computation. For each stock we select the last $L=180$ close-prices. Then for each specific time frequency we compute the realized variance, $RV$, of the exponentially weighted returns. Let $l$ be the time lag, $p$ the stock price, $w$ the exponential weights and $\rho$ the selected exponential decay speed, then for $l>1$:
\begin{eqnarray*}
\rho_l &=& \rho^{(L - 1) / (L - l)} \, , \\
w_i &=& (1 - \rho_l)(\rho_l)^i \, , \\
RV_l &=& \frac{1}{l} \sum_{i=l}^{L} w_i  \left( \log \left(\frac{p_{i}}{p_{i-l}} \right) \right)^{2}\, .
\end{eqnarray*}

Where $\rho_l = \rho^{(L - 1) / (L - l)}$ ensures that the sum of the exponential average weights is the same at each frequency $l$. For each stock we compute $RV_l$ for $l \in [1, 5]$ to compare the daily variance with the weekly one, where $l$ corresponds to the returns' frequency used for the computation. By re-weighting the variance by the time-scale we want to observe the spread with the commonly used theoretical variance rescaling. For example to obtain a weekly variance from a daily one it suffices to multiply it by $5$. This comes from the assumption that prices are i.i.d, see Diebold et al. (1997) \cite{fxd:1997ul} for a critique of this practice. Hence the expected behaviour would be for the function $RV_l$ to be constant with respect to the time scale $l$.

\subsection{Results for different stocks}
\begin{figure}[h]
\centering
\includegraphics[width = \figurewidth, height = \figureheight]{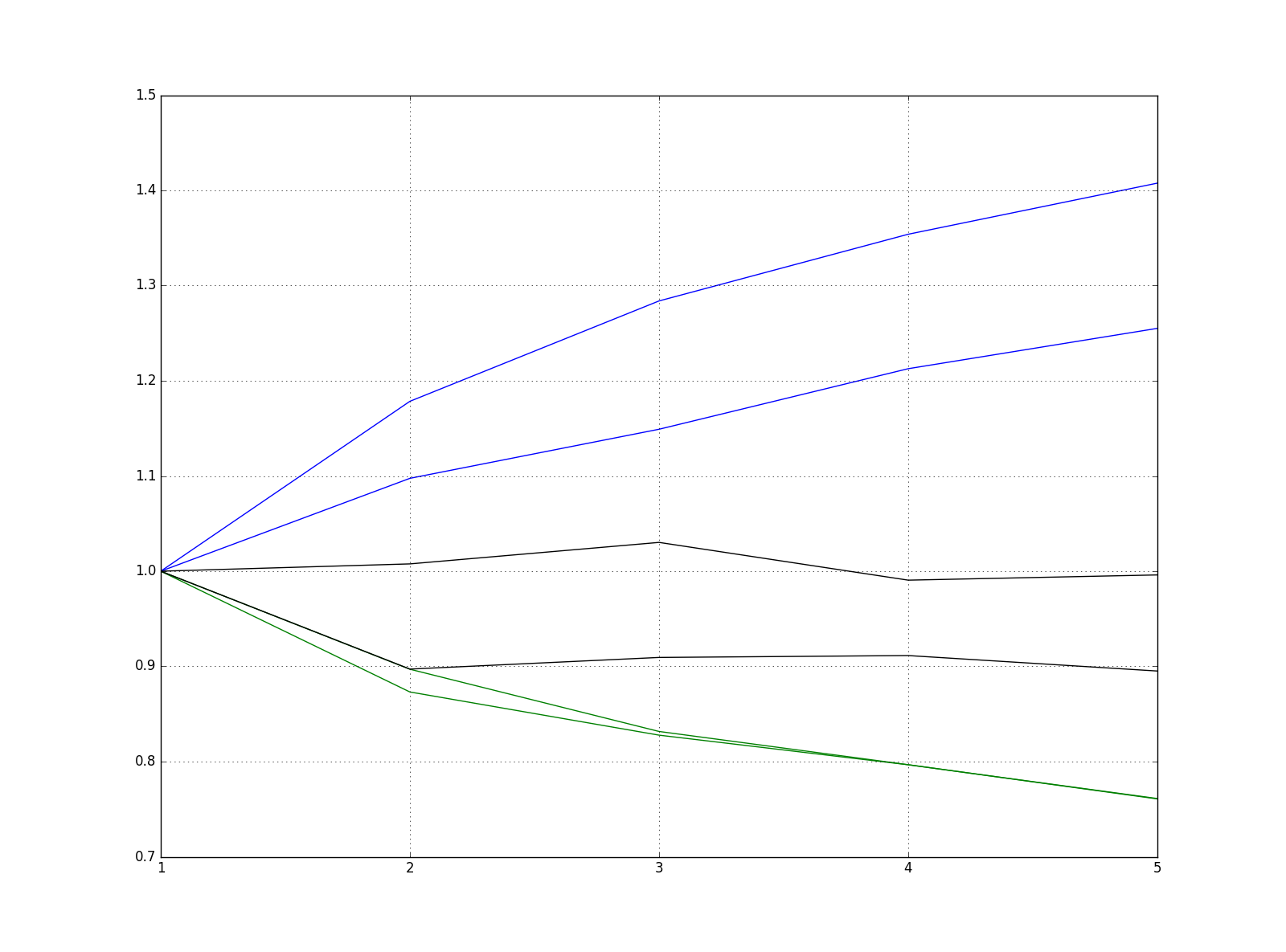}
\vspace{-20pt}
\caption{\label{fig:rvGamma} Variogram for $6$ different European stocks. The X-axis represents the time lag used to compute the realised variance while the Y-axis corresponds to the obtained RV values re-scaled to obtained a lag one value of $1$. The curves in black represent the expected behaviour. In blue are two selected stocks with increasing RV, while those in green are selected examples with decreasing RV.}
\end{figure}

In Figure \ref{fig:rvGamma} we show the result of the previously described computation for some selected stocks in the European market. This graphic exhibits three clusters: stocks with an increasing, constant, or decreasing variogram. Naturally the group of constant $RV_l$ are not strictly constant and should be better described as oscillating. The other two groups increase or decrease at different speeds. Indeed, while some seem to have an exponential variation, others are more linear. While the graphic shows only a few sampled stocks, we observed this behaviour across many different markets and stocks. Another interesting aspect of this phenomenon is that stocks with an increasing or decreasing variogram tend to keep this behaviour for some period of time.\bigskip

Following the description above, a stock with a bounding effect on its time series would have a decreasing variogram. On the other hand, stocks which tend to diverge will have a weekly variance higher than the daily one due to these sparse explosions, which creates an increasing variogram. While this explanation makes sense, it is hard to prove if this is the dominating effect. The interesting point here is that by simply observing the time series of prices at different frequencies we could potentially detect such behaviour. That leads us to go further and try to decompose the moves into different and interpretable economic factors that will explain the stock’s behaviour.

%%%%%%%%%%%%%%%%%%%%%%%%%%%%%%%%%%%%%%%%%%%%%%%%%%%%%%%%%%%%%%%%%%%%%%%%%%%%%%%%%%%%
\section{SGDLM model}
\label{sec:SGDLM}
\subsection{Description of the model}
The SGDLM model developed by \cite{Gruber:2016wn, Gruber:2016il, Zhao:2016kr, McAlinn:2016tm} brings together many different statistical techniques. Before it, the classic approach to working with DLMs in a multivariate setting was the inverse Wishart model; see \cite{Prado:2010wu} for a detailed description of this model and most tools used in the SGDLM. The motivation for this set-up was to study each time series of a multivariate model independently in parallel while still modelling the combined multivariate distribution. They used the variational Bayes approximation to approximate the multivariate model into independent individual distributions and then recouple them with importance sampling. This allows each time series' DLM to be different from the others. Let us denote by $y_{j,t}$ the observations, $\theta_{j,t}=(\phi_{j,t}, \gamma_{j,t})$ the state vector, $G_{j,t}$ the state evolution matrix, $F_{j,t}$ the vector of variables, $\nu_{j,t}$ the observations' noise independent of $\omega_{j,t}$ the states' noise. Therefore, $\lambda_{j,t}$ corresponds to the observations' precision and $W_{j,t}$ to the states' covariance matrix. Hence the DLM of series $j$ at time $t$ reads :
\begin{eqnarray*}
y_{j,t} &=& F_{jt} \theta_{j,t} + \nu_{j,t} = x_{j,t} \phi_{j,t} + y_{sp_t(j),t} \gamma_{j,t} + \nu_{j,t} \, ,  \\
\theta_{j,t} &=& G_{j,t} \theta_{j,t - 1} + \omega_{j,t} \, , \\
\text{with:} && \nu_{j,t} \sim \mathcal{N} (0,\lambda_{j,t}^{-1}) \, , \\
&& \omega_{j,t} \sim \mathcal{N}(0,W_{j,t})\, .
\end{eqnarray*}
Where $x_{j,t}$ and $\phi_{j,t}$ represent variables and coefficients of the internal predictors, while $ y_{sp_t(j),t}$ and $\gamma_{j,t} $ correspond to the variables and coefficients of the cross-series time varying conditional dependencies. In other words $(x_{j,t}, \phi_{j,t})$ represents the endogenous information and $(y_{sp_t(j),t}, \gamma_{j,t})$ the exogenous one. $sp_t(j)$ denotes the group of selected simultaneous parents which is adaptively revised over time to capture the changes in the market. Let us define $\mu_{j,t}=x_{j,t}\phi_{j,t}$ as the mean, $\Lambda_t$ as the precision-matrix with only $\lambda_{j,t}$ on the diagonal, and the matrix with the coefficients over the related parents by $\Gamma_t$, then we can write the multivariate model in matrix form:
\begin{eqnarray*}
y_t &\sim& \mathcal{N}\left( A_t \mu_t, \Sigma_t \right) \, , \\
\text{With:} && A_t = \left( I - \Gamma_t \right)^{-1} \, , \\
&& \Omega_t = \Sigma^{-1} = \left( I - \Gamma_t \right)^T \Lambda_t  \left( I - \Gamma_t \right) \, .
\end{eqnarray*}
Once this decomposition between endogenous and exogenous variables has been defined, the parents need to be chosen. In the first version of the model a fixed matrix defines the relationships between the different time series to select the exogenous variables. This proves particularly limiting as the performance of the whole model heavily depends on the choice of this matrix. In the latest evolution the authors added the two letters SG for Simultaneous Graphical. This means that a multivariate Wishart DLM runs in parallel to the main algorithm to build a relational graph and select which exogenous variables to include in the DLM. The selection of this group of parents will be detailed in the following section.\bigskip

It is interesting to emphasize how this set-up differs from the classic multivariate modelling approaches. Usually the multivariate distribution of the market will represent the cross-series relationships by non-zero elements in the covariance matrix, denoted $\Sigma$ here. This implicitly assumes the relationships to be symmetric. While in the SGDLM model the cross-series coefficients are present in each individual DLM as variables of the regression, $(y_{sp_t(j),t}, \gamma_{j,t})$, hence in the mean part of the individual DLM. Furthermore, in this model the step selecting the parents variable is  separated from the one computing its value. It first fits a multivariate inverse-Wishart distribution to the environment under study and use it to create a short-list of candidates. Members of this list are then included in the DLM formulation and the model will compute with its sequential update their coefficients. Therefore, it creates an asymmetrical relationship between the stocks in the environment; for each individual DLM the parents are selected because they impact the behaviour of the stock.\bigskip

By restricting each DLM to follow a log-normal-gamma distribution we can sequentially update the parameters of the DLM. This uses a Kalman filter with a Gamma distribution for the noise. Using this sequential update is possible thanks to the decomposition of the multivariate distribution into simple DLMs for which an extensive theory exists. Having independent equations for each time series will later allow us to include as endogenous factors an entire economic model. Another quality of this model is its capability to learn sequentially. Thanks to the Bayesian formulation and Kalman filter the model will continuously update the parameters of the distributions without depending on a specific look-back window. Indeed in the Kalman filter the past information is only present through the current value of the mean and variance of the distribution. That will allow us to follow the evolution of the coefficients through time and for the model to quickly adapt in case of rapid changes in the market. 

\subsection{Parents update}
\label{sec:sec_pa}
The group of parents for stock $j$, $sp_t(j)$, includes three subsets: the core $sp_{core,t}(j)$, up $sp_{up,t}(j)$ and down $sp_{down,t}(j)$ sets. The core parental set $sp_{core,t}(j)$ corresponds to the selected group of series used as predictors for $j$. In the up set are core-group candidates for entry into the parental set. In the down set are all the series previously in the core or up group.\bigskip

A multivariate Wishart DLM runs in parallel to the main algorithm to update the parental sets. With this model we obtain a dense covariance precision matrix which is used to select the candidates to the up set. For each series $j$ the $n_{max}$ series with the highest precision element become candidates for inclusion in the up set. As long as the size limit of the up set is not reached the $n_{max}$ series not currently in the up or core group are added to the up set. The candidates stay in this set during $\Delta T$ time before being considered for promotion to the core set. During that time they will be part of the parental set $y_{sp_t(i),t}$ used in the SGDLM. When the core group is full all the members of the up- and core-sets are ranked according to their signal to noise ratio, $a_t / R_t$, and the smallest values retired to the down set. Down set members will have their coefficients gradually put to zero over $\Delta T$ time periods.\bigskip
 
Issues arise in the choice for $n_{max}$, $\Delta T$ and the number of parents allowed. For example, if $n_{max}$ equals the number of parents allowed in the core and up sets, then the parents set can only be updated every $\Delta T$ times since it is full in-between. If the up-set does not have any size limit and the number of series considered for the up-set, $n_{max}$, is high then the number of parents could reach the total number of series $m$. These parameters will also influence the frequency at which the core-set can change all its members. A large $n_{max}$ will result in big and noisy up-sets and the bigger the up-set the higher the potential turnover rate of the core-set. A smaller $n_{max}$ will limit the maximum number of changes in the core-set, thus reducing the noise in those sets. If the number of allowed parents in the up-set is small, $\Delta T$ determines the turnover-frequency of the core-set. In addition, a short $\Delta T$ will put more importance in the choice of prior for the new members of the up-set, while, with a longer $\Delta T$ the prior choice becomes irrelevant in the signal to noise ratio used in the acceptance decision.\bigskip

These choices of parameters will influence the size of the matrices used in the following steps, since they increase as a square of the number of parents. Hence, while the choice of parameters does not influence the computation of the parents-update it does have a significant impact on the memory requirement of the other steps. The standard multivariate Wishart DLM and parent updates do not have any computationally intensive steps.

\subsection{Posterior update}
We update the posterior with the data measured at $\mathcal{D}_t$, and parameters computed at $t-1$. Since the time series are independent we can update them separately. Each independent DLM posterior distribution follows a Normal Gamma distribution $p(\theta_{j,t},\lambda_{j,t} | \mathcal{D}_t) \sim \mathcal{N} \mathcal{G}(m_{j,t}, C_{j,t}, n_{j,t}, s_{j,t})$:
\begin{eqnarray*}
\theta_{j,t} | \lambda_{j,t}, \mathcal{D}_t  &\sim& \mathcal{N}\left( m_{j,t}, C_{j,t} / \left( s_{j,t-1} \lambda_{j,t} \right) \right) \, , \\
\lambda_{j,t} | \mathcal{D}_t &\sim& \mathcal{G}\left(n_{j,t}/2, n_{j,t} s_{j,t-1}/2 \right) \, .
\end{eqnarray*}
Where $m_{j,t}$ corresponds to the mean, $n_{j,t}$ the shape of the Gamma distribution and $n_{j,t}s_{j,t-1}$ its scale. The notations were motivated by the observation that following those equations the distribution of $\theta_{j,t}$ is multivariate student-t distribution with $n_{j,t}$ degrees of freedom and scale matrix $C_{j,t}$. As a result of this simple formulation the posterior update follows the Kalman filter's equations. The multivariate posterior $P(\Theta_t, \Lambda_t | \mathcal{D}_t)$ is obtained by re-coupling those independent posterior distributions with importance sampling.\bigskip

The multivariate normal distribution sampling for $\left(\theta_{j,t} | \lambda_{j,t}, \mathcal{D}_t\right)$ will require a Cholesky decomposition of the covariance matrix. But since in practice the covariance matrix $C_{j,t}$ positive definiteness is not guaranteed we can run into numerical issues when sampling this variable. We therefore added in our implementation a shrinkage method to guarantee the positive definiteness of this covariance matrix estimation.\bigskip

This step quickly becomes computationally expensive with a large number of series, samples and allowed parents. Let $N_{MC}$ represent the number of Monte-Carlo samples, $m$ the number of series and $K$ the number of parents. Then the posterior sampling requires $m \times N_{MC}$ $K$-dimension multivariate normal-gamma computations.

\subsection{Recoupling with importance sampling}
The independent DLMs are recoupled to form the multivariate posterior distribution with importance sampling:
\begin{equation*}
p\left( \Theta_t, \Lambda_t | \mathcal{D}_t \right) \propto \left| I - \Gamma_t \right| \prod_{i=1}^{m} p(\theta_{j,t},\lambda_{j,t} | \mathcal{D}_t) \, .
\end{equation*}
Where $I$ is the identity matrix and $\left| - \right| $ corresponds to the determinant. Hence the importance sampling weights are defined by $ \alpha \propto \left| I - \Gamma_t \right| $. The weights are computed with Monte Carlo samples from the independent Normal-Gamma posterior distribution. Since the number of parents will stay low compare to $m$ the coefficient matrix $\Gamma_t$ will be sparse. Following previous section reasoning, the exact posterior requires the computation of $N_{MC}$ $m \times m$-sparse-matrices determinants to obtain the importance sampling weights.\bigskip

In \cite{Gruber:2016il} they used the importance sampling entropy as a proxy of the Kullback-Leibler distance and thus of the quality of the model-fit. They define the entropy $H$ by:
\begin{equation*}
H_N = \sum_{n=1}^N \alpha_n \log ( N \alpha_n ) \, .
\end{equation*}
We will study this variable as a proxy of the environment stress. The higher the entropy the lower the approximation quality, which may be interpreted as a sign of increased environment-variance as a result of divergence between the model and the observations. 

\subsection{Decoupling with variational Bayes}
The one-step-ahead inference corresponds to the prior update $p\left( \Theta_{t+1}, \Lambda_{t+1} | \mathcal{I}_{t+1} \mathcal{D}_t \right)$. The authors used Variational Bayes to decouple the multivariate distribution into independent series and thus process the time series in parallel. The Variational Bayes decomposition approximates the multivariate posterior  $p\left( \Theta_t, \Lambda_t | \mathcal{D}_t \right)$ by a product of Normal-Gammas $p(\theta_{j,t}, \lambda_{j,t} | \mathcal{D}_t) \sim \mathcal{N}\mathcal{G}(m_{j,t}, C_{j,t}, n_{j,t}, s_{j,t})$ by minimizing the Kullback-Leibler (KL) distance in the usual manner.\bigskip  

The resulting KL distance at each time $t$ gives information on the quality of the fit. Significant changes in the environment under study would decrease the approximation quality and thus increase the KL distance. \cite{Gruber:2016wn} applied this model on the S\&P$500$ time series and observed similarities between a rescaled KL measure and the St. Louis Federal Reserve Bank Financial Stress Index. That observation led them to study further the strength of this relation, see Section \ref{sec:results_risk} for a more detailed discussion.\bigskip

Once the multivariate distribution is decomposed into independent DLMs the parameters are updated independently in parallel to $t+1$. The updated parental set updates the filter's parameters $(a_{j,t+1}, R_{j,t+1}, r_{j,t+1}, W_{j,t+1})$. The innovations follow a Gamma-Beta Stochastic volatility update $\lambda_{j,t+1} = \lambda_{j,t} b_{j,t} / \beta_{\lambda}$ where $b_{j,t} \sim \text{Beta} \left( \beta_{\lambda} n_{j,t} / 2, (1-\beta_{\lambda}) n_{j,t} / 2\right)$. Where the parameter $\beta_{\lambda}$ influences the smoothness of the predicted $\lambda$. The Gamma-Beta Stochastic volatilities appears through the gamma parameter update $r_{j,t+1} = \beta_{\lambda} n_{j,t}$. According to the DLM equations, the states follow the update $\theta_{j,t+1} = G_{j,t+1} \theta_{j, t} + \omega_{j,t+1}$ with $\omega_{j,t+1} \sim \mathcal{N}(0, W_{j,t+1}/(s_{j,t} \lambda_{j,t+1}))$.

\subsection{Inference}
Each DLM has an updated prior distribution $p(\theta_{j,t+1}, \lambda_{j,t+1} | \mathcal{D}_t) \sim \mathcal{N}\mathcal{G}(a_{j,t+1}, R_{j,t+1}, r_{j,t+1}, s_{j,t})$. Samples from these independent priors are combined to obtain the multivariate distribution $y_{t+1} \sim \mathcal{N}(A_{t+1} \mu_{t+1},  \Sigma_{t+1})$, with first moments, $A_{t+1} \mu_{t+1}$, and second moments, $\Sigma_{t+1}$. This step is computationally expensive and requires the computation of $N_{MC} \times m$ multivariate normal gamma distribution to obtain $p(\theta_{j,t+1}, \lambda_{j,t+1} | \mathcal{D}_t)$, $N_{MC}$ inverses and matrix-matrix products of $(m, m)$ matrices to compute the first two moments $A_{t+1}\mu_{t+1} = \left( I - \Gamma_{t+1} \right)^{-1} \mu_{t+1}$ and $\Sigma_{t+1}= \left( \left( I - \Gamma_{t+1} \right)^T \Lambda_{t+1} \left( I - \Gamma_{t+1} \right) \right)^{-1} $.\bigskip

Inference more than one day ahead can be computed recursively with the previous one-step-ahead prediction. Before the next step prediction, the filter's parameters $a_{j,t+1}$ and $R_{j,t+1}$ are updated with the previously inferred value. As before we produce Monte-Carlo samples of the first two moments to infer time $t+k$ with the computed parameters at $t$ and inferred values at $t + k - 1$. The difficulty comes from the internal predictors variables $x_{j,t}$ which requires to have the values for the next step update. Hence it is only possible to recursively produce inference more than one step ahead if the predictors' variables can be updated with the previously inferred values.

\subsection{Parameters sensitivity}
Due to its Bayesian structure the model has a few hyper-parameters. For the parent step the size of the different groups is fixed and must be decided with the update frequency. Due to the impact of these variables  on the memory requirement, the hardware limits the possible range. Regarding the parent set, the newly selected variables will need a mean and variance to initialise their distribution. We choose a zero mean and small variance for them not to impact the rest of the regression knowing that this value will quickly be updated with the Kalman filter in the subsequent steps.\bigskip

The states $\theta_{j,t}$ follow a random walk update $\theta_{j, t+1} = G_{j, t+1} \theta_{j, t} + \omega_{j, t}$, where the covariance matrix for the noise terms $W_{j, t}$ follow a block discounting update; we refer to the appendix for the detailed equations. This step uses discounting factors $\delta_{j, \phi}$ for the endogenous variables and $\delta_{j, \gamma}$ for the parental ones. The multivariate Wishart also uses this discounted covariance update technique with parameter $\delta_w$. These parameters influence the variance-of-variance of the states coefficients. Similarly, the precision $\lambda_j$ follows a Gamma-Beta Stochastic volatility update which needs a parameter $\beta_{\lambda}$ that influences the variance of this update. The same parameter, $\beta_w$ is also present with the same role in the multivariate Wishart. These are the five hyper-parameters of the SGDLM framework. \cite{Gruber:2016il} studied the influence of the discounting parameters and gave theoretical bounds for the betas.

%%%%%%%%%%%%%%%%%%%%%%%%%%%%%%%%%%%%%%%%%%%%%%%%%%%%%%%%%%%%%%%%%%%%%%%%%%%%%%%%%%%%
\section{The Heterogeneous AR model explained}
\label{sec:HAR}
\subsection{The motivations for the model}
\cite{Corsi:2004eg, Corsi:2009ue, Corsi:2008ei} developed the HAR-RV, Heterogeneous Autoregressive Realised Volatility model for two main reasons: first the model must reproduce the fat-tails, long-memory, scaling and multifractal volatility behaviours; second it must be easy to estimate and keep clear economic interpretations. They used the easy to compute Realised Volatility as a proxy of the unobservable latent volatility. Additive models were known for their clear economic interpretations and easy calibration but not known to reproduce multifractality. That was before \cite{LeBaron:2001cp} reproduced multifractality and long-memory with a simple three factor model where the variables represented three different and specifically chosen frequencies. Hence, a simple additive model with correctly chosen frequencies can reproduce the expected stylised facts. This led to the HAR model with the three, short, medium and long frequencies defined as: day, week and month.\bigskip

Different papers studied the asymmetric propagation of the volatilities between different time scales. The HARCH model introduced by Muller,  \cite{Muller:1997he, Dacorogna:1997ez} followed the heterogeneous market hypothesis to decompose the volatility into different time-scale dependent components to prove this asymmetry. An interpretation behind this effect is the heterogeneous objectives of markets participants, one of them being the investment horizon. In other words, the daily volatilities impact short-term traders more than long-term ones while low frequency movements impact both. A mathematical representation of this idea creates an additive cascade of the different market components. \cite{Corsi:2004eg} build the cascade with an $AR(1)$ models of the RV at each frequency with a coefficient to relate it to the closest lower frequency component.\bigskip

This model reproduces the long-memory and multifractality behaviours without being in the long-memory model class and keeps an economic interpretation. In their paper, they studied the coefficients to learn which frequency is driving the volatility move. They use the observation that the low frequency coefficients are non-negligible to explain the daily volatility moves as an argument for the volatility cascade.

\subsection{The original HAR-RV}
The model recreates the volatility cascade by incorporating dependencies on lower scale volatility components at each frequency. Thus the return process is a function of the cascade's highest frequency component, which is the daily volatility $\sigma_t^d$ here. Let us consider the returns $r_t$ to follow:
\begin{equation*}
r_t = \sigma_t^d \epsilon_t \, , 
\end{equation*}
with Gaussian noise $\epsilon_t \sim \mathcal{N}(0, 1)$. Let $RV_t^s$ represents the observed Realised Volatility at $t$ for time scale $s$. They define the daily RV by:
\begin{equation*}
RV_t^d = \sqrt{\sum_{j=0}^{M-1} r^2_{t-j\Delta}} \, , 
\end{equation*}
with $\Delta$ the sampling frequency and $M$ the resulting number of points. By doing so they obtain a daily volatility computed from high frequency returns. The lower-frequency RVs are built by averaging the daily ones. e.g. the weekly RV is:
\begin{equation*}
RV_t^w=\left( RV_{t-1d}^d + RV_{t-2d}^d  + \dots + RV_{t-5d}^d\right) / 5 \, , 
\end{equation*}
Following the volatility cascade idea, at each time scale the unobserved partial volatility process  $\tilde{\sigma}_t^s$ has an $AR(1)$ structure with a coefficient for the expected volatility at the next lower scale. If we consider three frequencies, $d$ daily, $w$ weekly, $m$ monthly the cascade model reads:
\begin{eqnarray*}
\tilde{\sigma}_{t + 1m}^m &=& c^m + \phi^m RV_t^m + \tilde{\omega}_{t + 1m}^m \, , \\
\tilde{\sigma}_{t + 1w}^w &=& c^w + \phi^w RV_t^w + \gamma^w E_t[ \tilde{\sigma}_{t + 1m}^m] + \tilde{\omega}_{t + 1w}^w \, , \\
\tilde{\sigma}_{t + 1d}^d &=& c^d + \phi^d RV_t^d + \gamma^d E_t[ \tilde{\sigma}_{t + 1d}^d] + \tilde{\omega}_{t + 1d}^d \, , 
\end{eqnarray*}
where $c^s$ corresponds to a constant, $phi^s$ the coefficient of the RV variable and $\gamma^s$ the coefficient representing the dependence on the closest lower scale. The noises $\tilde{\omega}_{t}^s$ are independent in time and between each others. By recursion we obtain:
\begin{equation*}
\tilde{\sigma}_{t + 1d}^d = c + \beta^d RV_t^d +\beta^w RV_t^w + \beta^m RV_t^m + \tilde{\omega}_{t+1d}^d \, , 
\end{equation*}
Where $\beta^s$ represent the different coefficients. With the relation between the daily latent volatility measure $\tilde{\sigma}_{t + 1d}^d$ and its estimate $RV_{t+1d}^d$ written as $\tilde{\sigma}_{t + 1d}^d = RV_{t+1d}^d + \omega_{t+1d}^d$, we obtain the following time series representation of the cascade model:
\begin{equation*}
RV_{t + 1d}^d = c + \beta^d RV_t^d +\beta^w RV_t^w + \beta^m RV_t^m + \omega_{t+1d} \, .
\end{equation*}
Where $\omega_{t+1d} = \tilde{\omega}_{t+1d}^d - \omega_{t+1d}^d$.  Hence the variance is modelled as a combination of different frequency components. Due to its simple additive form the authors suggested the use of classic least square regression to compute the parameters. In their paper they used all the available history as an increasing window on which to fit the model. Hence the first point uses only $30$ days while the last one uses the whole dataset. In their original paper they also discussed using a moving window regression to obtain a time series evolution of the weight but did not show any results.

\subsection{The extended version for leverage and volatility-of-volatility}
In order to avoid negativity issues they used the logarithm of the realized-variance $log(RV_t^l)$. They also extended the original HAR-RV to model the leverage effect by adding variables at each frequency representing past positive $r_s^+$ and negative $r_s^-$ returns, with $s$ the time scale. The cascade model with leverage coefficients $\gamma$ for each time scale becomes:
\begin{eqnarray*}
log\left( RV_{t + 1d}^d \right) = c &+& \beta^d log\left(RV_t^d\right) + \gamma^{(d)+} r_t^{(d)+} + \gamma^{(d)-} r_t^{(d)-} \, , \\
&+& \beta^w log\left(RV_t^w\right) + \gamma^{(w)+} r_t^{(w)+} + \gamma^{(w)-} r_t^{(w)-} \, , \\
&+& \beta^m log\left(RV_t^m\right) + \gamma^{(m)+} r_t^{(m)+} + \gamma^{(m)-} r_t^{(m)-} \, ,\\
&+& \omega_{t+1d} \, .
\end{eqnarray*}
In \cite{Corsi:2008ei} they presented a HAR-RV model with Normal Inverse Gamma variance to model the variance-of-variance. While this model did not out-perform alternative models for all of the tests, it did prove to be the best regarding forecasts of the distribution of the realized-variance. Moreover they showed that incorporating time variation of the variance of RV, without necessarily using an inverse gamma distribution, produced better results for inference and for the fit of the RV distribution.

%%%%%%%%%%%%%%%%%%%%%%%%%%%%%%%%%%%%%%%%%%%%%%%%%%%%%%%%%%%%%%%%%%%%%%%%%%%%%%%%%%%%
\section{A complete model of the market}
\label{sec:prop}
\subsection{The motivations for the model}
The HAR-RV model of \cite{Corsi:2004eg, Corsi:2009ue, Corsi:2008ei} allowed them to identify which frequency, and thus which traders were influencing the stock. On the other-hand the SGDLM framework of \cite{Gruber:2016wn, Gruber:2016il, Zhao:2016kr, McAlinn:2016tm} allowed them to dynamically create a graph of connections between stocks in a market and incorporate this information into a multivariate distribution of the environment. Combining those two models creates a time varying DLM for each asset with factors representing the influence of different frequencies to identify which trader is driving the moves and others for the influence of selected stocks in the market. Furthermore, \cite{Corsi:2008ei} studied the importance of including a time varying noise component and more specifically a Normal inverse gamma distribution to efficiently model the variance-of-variance distribution. The SGDLM model uses exactly that distribution for the noise terms since it is the conjugate distribution of a Normal with unknown variance. In addition, both models follow a simple additive set-up, hence the extension of the classic HAR-RV to incorporate leverage and jumps is straightforward to include into the DLM.\bigskip

Let us recall the initial variogram study. The study of the evolution of the RV while changing the returns' scale from daily to weekly allowed us to detect abnormal patterns in the time series, which were potentially related to option hedging. The combination of the HAR-RV and SGDLM models allow us to go further into this decomposition. Each time series is decomposed into two groups of factors. The exogenous factors consist of the influence of the environment on the series, and this group combines the variables selected during the parent selection phase. Secondly, the endogenous factors represent the information internal to the time series. This is composed of the previous RV and leverage variables for different frequencies.\bigskip

With this decomposition we aim to understand whether the moves are being driven by external or internal factors, and also at which frequency. In addition, it is interesting to study the evolution of the spread between endogenous and exogenous groups. The intuition here is that a company with a variance being pushed mainly by external factors should behave differently from one which mainly follows endogenous ones. Looking at a different scale, can we relate different market situations to phases where the stocks are being mainly driven by exogenous factors, endogenous ones or that the spread between these two is increasing. The results presented in sections \ref{sec:results} \& \ref{sec:resultsDiffEnv} provide an answer of these questions.

\subsection{Extending the standard DLM}
While \cite{Corsi:2004eg, Corsi:2009ue, Corsi:2008ei} used high frequency intra-day data to obtain a daily-variance, in this study we only make use of of end-of-day prices. At this low frequency we define the $s$-scale RV as the sum of squared $s$-scale log-returns, $r_{j,k}^s=\sum_{k=t-i}^{t - i - s} r_{j,k}$, over a time window $L$. With a weighting kernel $w$, the $s$-scale-RV is defined by:
\begin{eqnarray*}
RV_{j, t}^s &=& \frac{1}{s} \sum_{i=0}^{L - 1} w_i \left( \sum_{k=t-i}^{t - i - s} r_{j,k} \right)^2 \, .
\end{eqnarray*}
We will use an exponential moving average weighting kernel for $w$. To reproduce the HAR-RV three scales framework we will use daily $(s=1)$, weekly $(s=5)$ and monthly $(s=20)$ frequencies. If we suppose unknown RV precision $\eta_{j, t}$ as in the original SGDLM model, the resulting $RV_{j, t}^d$ is Normal-inverse-Gamma. Following \cite{Corsi:2008ei} we use the logarithm of the RV to avoid negativity issues, hence obtaining a log-Normal-Inverse-Gamma distribution for each time series. \bigskip

A DLM is defined by a distribution on the output variable and another on the state variables. Following the notations introduced for the SGDLM, we can write the HAR-RV cascade model of one stock $j$ as:
\begin{eqnarray*}
RV^d_{j,t} &=& F_{jt} \theta_{j,t} + \nu_{j,t} \, , \\
\theta_{j,t} &=& G_{j,t} \theta_{j,t - 1} + \omega_{j,t} \, , \\
\text{with:} && \nu_{j,t} \sim \mathcal{N} (0,\lambda_{j,t}^{-1}) \, , \\
&& \omega_{j,t} \sim \mathcal{N}(0,W_{j,t}) \, .
\end{eqnarray*}
Where $\theta_{j, t}$ is the state vector and $F_{j,t}$ the vector of variables. The multivariate state-noise is $W_{t} \sim \mathcal{N}(0, V_t)$ with $V_t$ the covariance matrix. The evolution matrix $G_{j, t}$ is diagonal with non-zero elements for the selected parents. Following the SGDLM approach we can extend this DLM to include $K$ parents from the $m$ stocks in the environment under study. The state vector becomes:
\begin{equation*}
\theta_{j,t} =  \left( c_{j, t}, \beta_{j, t}^d, \beta_{j, t}^w, \beta_{j, t}^m, \gamma_{1, t}, \dots,  \gamma_{K, t} \right)^T \, , 
\end{equation*}
Where the $\beta$ represent the coefficients of the endogenous variables, here the offset and three frequencies from the HAR-RV model, and the $\gamma$ the coefficients of the exogeneous ones. Hence the corresponding vector of variables is:
\begin{equation*}
F_{j, t} = \left( 1, RV_{j, t}^d, RV_{j, t}^w, RV_{j, t}^m, RV_{sp_t(1), t}^d, \dots, RV_{sp_{t}(K), t}^d \right) \, .
\end{equation*}
Extending this set-up to include the leverage effect is straightforward. The complete model will therefore be composed of internal coefficients, the realised variance and leverage effect coefficients at different frequencies, and external ones, the group of daily realised variance of the selected parents. While for the HAR-RV model they considered the variance at different scales to be the average of the daily one over different time windows we considered it to be the RV over the same time window but with different frequencies of returns. We will refer to this combined model as H-SGDLM.
% Since we do not know a priori what the most explanatory variable is, we use a model with nine RV variables corresponding to the RV computed for the three time windows and three return frequencies: daily, weekly and monthly. We will refer to this combined model as H-SGDLM.

\subsection{Add OHLC data}
In this section we will a present an example of a possible extension of the H-SGDLM to incorporate new endogenous variables. As a result of the simple additive structure of the H-SGDLM it is easy to extend as long as the constraint of having Normal distributions is respected. A day of trading is often summarized with the Open, High, Low, Close (OHLC) data. Open and Close correspond to the first and last traded price of the day while, High and Low are the highest and lowest traded price of the day. We studied the correlation of different metrics combining those variables and the underlying time series of close prices. Let us note a stock price as $S$ then the variables with the highest correlation are:
\begin{eqnarray*}
r^{low}_t &=& log(S^{low}_t) - log(S^{low}_{t-1}) \, , \\
CH_t &=& \frac{S^{high}_t - S^{close}_t }{S^{high}_t - S^{low}_t} - 0.5 \, , \\
COHL_t &=& \frac{S^{close}_t - S^{open}_t }{S^{high}_t - S^{low}_t} \, .
\end{eqnarray*}
Following the idea of incorporating the OHLC data to make the model more responsive to recent moves we can add them to the vector of endogenous variable. Instead of increasing its side we deleted the less informative variables from the previous formulation which corresponded to the RV variables computed of a time window of a week. Thus we obtain:
\begin{eqnarray*}
H_{j, t} &=& (1, RV_{j, t}^d, RV_{j, t}^w, RV_{j, t}^m,\\
&& (r_{j, t}^d)^2, (r_{j, t}^w)^2, (r_{j, t}^m)^2,\\
&& r^{low}_{j,t}, CH_{j,t}, COHL_{j,t},\\
&&  r_{j,t}^{(d)+}, r_{j,t}^{(d)-}, r_{j,t}^{(w)+}, r_{j,t}^{(w)-}, r_{j,t}^{(m)+}, r_{j,t}^{(m)-} )^T \, .
\end{eqnarray*}
In our simulations, incorporating OHLC variables did not improve the performance of the H-SGDLM when applied to model the log-RV. But when in section \ref{sec:results_price} we adapted the H-SGDLM to predict the logarithm of the prices on different environment the OHLC data improved the results.

\subsection{The implementation on a GPU}
Since the SGDLM framework decomposes the multivariate distribution into independent DLMs they can be updated in parallel. \cite{Gruber:2016il} used a GPU implementation developed in C++ with the CUDA library from Nvidia and ran it on a cluster of GPUs. We used a simpler approach in the form of the TensorFlow library. This library developed by Google is primarily made for neural networks and deep learning architecture. Nevertheless, the wide array of functions available are all optimised to process large tensors on GPU; hence, we used TensorFlow as a library for tensor computation on GPU and implemented the H-SGDLM with it.\bigskip

The two main implementation issues are firstly, the matrix inversion for the Kalman update, and secondly the size of the matrix of the full multivariate distribution. For the matrix inverse we added a regularization step to guaranty its positive definiteness before inversion and monitor its values to avoid any divergence. In addition, the scale of the variables used in the model also has an impact, which is why we rescale them. For the memory issue we used the fact that the coefficient matrix $\Theta_t$, the update matrix $G_t$ and the covariance matrices are all sparse. Unfortunately, at the time of writing the TensorFlow library did not have the necessary functions to perform the sparse matrix manipulations we needed. Hence, we stored a dense version of those matrices and a matrix of indices to rebuild them when needed for computation. While this is not optimal, it suffices for the purpose of testing our model.\bigskip

Due to the memory consumption of these matrices and the limited memory available on a single GPU some parameters had to be bounded. More specifically, the number of allowed parents into the core, up and down sets and the number of time step between the update of those groups. The GPU memory places limits on the feasible number of Monte-Carlo samples too. One option would be to store a set of samples on the permanent memory and continue sampling on the GPU, however the added memory transfer counterbalances the gain in GPU processing. Since we only had one GPU at our disposal we could not reproduce the approach taken in \cite{Gruber:2016il} of distributing the sampling on multiple GPUs to allow for larger sets. 

%%%%%%%%%%%%%%%%%%%%%%%%%%%%%%%%%%%%%%%%%%%%%%%%%%%%%%%%%%%%%%%%%%%%%%%%%%%%%%%%%%%%
\section{Results}
\label{sec:results}
\subsection{On a simulated environment}
Before studying real data we tested the algorithm in a simulated environment. Each stock prices is assumed to follow a random walk with a Brownian motion with different means and variances which are distributed randomly but constant through time. With this process we created a dataset of $300$ time series. The parent selection using the multivariate Wishart does not show any clear clear pattern; the connections are noisy and seem random. This behaviour is also reflected in the exogenous coefficients, their absolute value stays low throughout with no clear patterns. All of these observations meet the key objective; namely that given a random environment with no connections between the time series, indeed the model does not find any. For the endogenous variables, the autoregressive coefficient of the RV at the previous time point stands out from the rest. Again, this was expected since the simulated time series follow a random walk.

\subsection{Description of the dataset}
% In this section we will only work with data from the S\&P$500$. We selected all the stocks that have a history of prices extending to year $2000$ which resulted in a set of $m=378$ stocks. The aim of this selection was mainly to reduce the number of time series since we were limited by the GPU memory. Due to the model set-up the bigger the environment under study the better should the results be. Indeed, we see that the results in Section \ref{sec:resultsEU} for a bigger environment with $500$ time series are better. For the exponential averages we used a parameter $\rho=0.98$ and a time window of $rv_L = 40$ days. For any missing $log(RV)$ point we put a value of $0$. Since we initiate the variables randomly the model needs some time to converge, so all the graphics and data shown below are between $10/04/2001$ and $03/06/2019$. \bigskip
In this section we use data from European stocks. The data corresponds to the $487$ most liquid stocks selected from thousands of stocks in different European markets. We start the backtest from $2000$ but since not all stocks have such a long trading history we assume $log(RV)$ to be zero when the price is missing. With that number of time series the available GPU memory only allows us to work $500$ Monte Carlo samples. For the exponential averages we used a parameter $\rho=0.98$ and a time window of $rv_L = 40$ days. Since we initiate the variables randomly the model needs some time to converge, so all the graphics and data shown below are between $01/06/2001$ and $04/06/2019$.\bigskip

Regarding the update of the parental sets, the sizes are for the core set $nCore=5$, for the up and down sets $nUp=nDown=5$. With a time interval of $\Delta T = 10$ the core set will need a minimum of $10$ days to be completely updated. These numbers were also chosen to obtain a number of parents' variables less than $15$. The objective is to have a balance between the number of endogenous and exogenous variables, as well as a low percentage of parents relative to the environment. In this section we compare the performance of our proposed algorithm with the previously published HAR-RV and SGDLM models. Since they did not use OHLC variables we won’t either. In the following section \ref{sec:resultsDiffEnv} we apply the model on different investment environments.

% \subsection{Test of convergence}
% \textcolor{red}{Need to add study of the intercept value, and empirical distribution of the noise to check it is white noise and that KF behaves optimally}

\subsection{Direct inference}
\label{sec:results_met_EU}
This section compares the one step ahead inference computed from the proposed model to those from the HAR-RV and SGDLM models. Our model was developed to study the evolution of the coefficients to learn which one is driving the market, hence the quality of the one step ahead inference is used to assess the accuracy of this decomposition. We ran the algorithm on daily $log(RV)$ which produced one day ahead predictions. To put the metrics in perspective it is interesting to realize that the average daily return  of $log(RV)$ over the whole dataset is $0.0$, while its standard deviation is $0.28$. Therefore, simply predicting a constant value for the next day already produces a good mean absolute deviation (MAD), hence we consider this case as model $t-1$ in table \ref{tab:mad_EU}. \cite{Corsi:2004eg, Corsi:2009ue, Corsi:2008ei} assessed the out of sample performance with the $R^2$ of Minor-Zarnowitz, Root Mean Squared Error (RMSE) and Mean Absolute Deviation (MAD). West et al. used the MAD to assess the one point ahead forecast and the performance of a portfolio optimisation to assess the quality of the multivariate model and especially the resulting sparse covariance matrix. We ran on our dataset the HAR-RV model with leverage effect and the SGDLM model with only a constant and the previous value as endogenous variables. All the metrics in table \ref{tab:mad_EU} correspond to averages over the individual ones obtained for each time series. The similarity between the HAR-RV model and $t-1$ model can be easily explained: when we look at the values of the different coefficients in the HAR decomposition, the previous $log(RV_{t-1})$ variable has a weight close to $1$.\bigskip

When assessing inference of a Bayesian model, the median absolute deviation between the predicted mean of the distribution and the target time series does not take into account the quality of the inferred distribution. A more interesting metric can be computed by using the predicted confidence interval and the percentage of measured points which occurred inside it. The better the forecast, the smaller the confidence interval and the higher the percentage of points in the predicted interval should be. Since the HAR-RV employed a classic least-square parametrization, we used a multiple of the squared standard deviation of the predictions over the past $30$ points to compute the confidence interval. In order to compare the performance of the three models, HAR-RV, SGDLM and ours, we used a multiple of the standard deviations to obtain similar sizes of confidence interval and thus compare the percentage of correct predictions for that interval. Regarding the choice of the interval, we selected a size smaller than the twice the average move size, i.e. for an average move size of $5\%$ we select a confidence interval lower but close to $10\%$. To compare the three models we select an average confidence interval of similar size so we can compare the predicted move direction and potential confidence interval increase at the jumps. The perfect model should have very low confidence interval on average with large increase when a jump is predicted.\bigskip

Since the motivation behind combining endogenous information representing the volatility cascade and exogenous information representing the influence of the market was to predict the underlying risk, we expect our model to perform well in predicting large moves. Hence, Table \ref{tab:confIntabs_EU} shows the percentage metric for moves with an absolute return larger than a certain threshold. But, since $68,7\%$ of the moves of $log(RV)$ in our dataset are negative the quality of the model is assessed by the prediction of increases in variance. Therefore we also show in Table \ref{tab:confInt_EU} the percentage of correct predictions when focusing on volatility increases above a certain threshold. In every scenarios our model correctly predicted more than $62.0\%$ of the moves, positive or negative. While it correctly predicted $63.89\%$ of moves larger than $9.28\%$ and $63.95\%$ of the increases higher than $10.04\%$. On the same data, for the same thresholds, the classic HAR-RV and SGDLM models never achieves more than $54.14\%$ and $40.41\%$ respectively. Our combined model behaves as expected and outperforms both previous approaches for positive, negative, small and large moves. Interestingly, the bigger the moves the better the performance of our model and the bigger the gap with the other two models.\bigskip
% In this case of large upward moves our model correctly predicted $63.8\%$ of the jumps of average size $9.9\%$, and $60.0\%$ of the increases of average size $4.0\%$. These figures represent  a considerable increase compared to the other two approaches. Indeed, while the HAR-RV model correctly predicts more than half the moves, its performance decreases with the size of the returns to $29.8\%$ for moves higher than $9.9\%$. On the other hand, the SGDLM correctly predicts $59.6\%$ of the large ones. The combine model behaves as expected and outperforms both previous approaches for both small and large moves. Interestingly, the bigger the moves the better the performance of our model and the higher the difference with the other two models.\bigskip

\begin{table}[h]
\centering
\caption{Performance metrics of the one-day-ahead inference obtained for $487$ European stocks.}

\begin{subtable}{\textwidth}
\centering
\begin{tabular}{ l | c | c | c | c}
\multicolumn{1}{c | }{Model} & t - 1 & HAR-RV & SGDLM & Full \\
\hline \hline
Median ADV         & $0.010$  & $0.012$  & $0.015$ &  $0.013$\\ 
Median RMSE      & $0.038$  & $0.040$  & $0.050$ & $0.043$\\
Median $R^2$     & $0.990$  & $0.989$  & $0.985$   &  $0.988$\\
 M-Z coefficient  & N/A          & $0.977$ & $1.006$  &  $0.991$\\
\end{tabular}
\caption{\label{tab:mad_EU} One day ahead inference comparison between different models. Median ADV corresponds to the Median Absolute deviation. Median RMSE is the median of the Root Mean Squared Errors. M-Z coefficient corresponds to the Minor-Zarnowitz regression coefficient, i.e. the regression of the observed values on the predicted ones.}
\end{subtable}
\end{table}

\begin{table}[p]
\centering
\caption{Performance metrics with confidence interval of one-day-ahead inference obtained for $487$ European stocks.}

\begin{subtable}{\textwidth}
\centering
\begin{tabular}{ l | l | c | c | c }
\multicolumn{2}{c | }{Model} & HAR-RV & SGDLM & Full \\
\hline \hline
\multirow{2}{*}{without threshold}  & Interval size   & $3.96\%$ & $3.97\%$ & $3.95\%$ \\ 
  								&  \% in  & $64.55\%$ & $49.43\%$ & $62.52\%$ \\
\hline
\multirow{2}{*}{with  mean$(| r |) = 5.73\%$ } &  Interval size  & $10.75\%$ & $10.74\%$ & $10.71\%$ \\
  									      &  \% in   & $55.41\%$ & $37,43\%$ &  $62.38\%$ \\
\hline
\multirow{2}{*}{with mean$(| r |) = 7.55\%$ }  &  Interval size  & $14.22\%$ & $14.20\%$ &  $14.20\%$ \\
  									     &  \% in   & $53.43\%$ & $34.5\%$ &  $62.33\%$ \\
\hline
\multirow{2}{*}{with mean$(| r |) = 9.28\%$ }  &  Interval size  & $18.02\%$ & $18.04\%$ &  $18.06$ \\
  									      &  \% in   & $53.24\%$ & $34.69\%$ &  $63.89\%$ \\
\end{tabular}
\caption{\label{tab:confIntabs_EU}In order to compare the different models, we fixed a confidence interval size but allowed for different means in the predicted distribution. This table shows the percentage of measured $log(RV)$ which were in the predicted confidence interval for the selected absolute-returns above a certain threshold. The first columns shows the average absolute-return,  mean$(| r |)$, of those selected points. The confidence intervals are selected to be smaller than twice the average-return of the selected points, here mean$(| r |)$. This table shows the ability of our model to predict large, positive and negative, moves of any size.}
\end{subtable}

\bigskip

\begin{subtable}{\textwidth}
\centering
\begin{tabular}{ l | l | c | c | c }
\multicolumn{2}{c | }{Model} & HAR-RV & SGDLM & Full \\
\hline \hline
\multirow{2}{*}{with  mean$ (r ) = 6.92\%$ } &  Interval size  & $13.06\%$ & $13.11\%$ & $13.10\%$ \\
  									    &  \% in   & $56.30\%$ & $43.75\%$ &  $62.69\%$ \\
\hline
\multirow{2}{*}{with mean$ (r ) = 8.48\%$ }  &  Interval size  & $16.63\%$ & $16.71\%$ &  $16.63\%$ \\
  									    &  \% in   & $55.80\%$ & $43.30\%$ &  $64.12\%$ \\
\hline
\multirow{2}{*}{with mean$ (r)  = 10.04\%$ }  &  Interval size  & $19.71\%$ & $19.71\%$ &  $19.73\%$ \\
 									    &  \% in   & $54.14\%$ & $40.41\%$ &  $63.95\%$ \\
\end{tabular}
\caption{\label{tab:confInt_EU}In order to compare the different models, we fixed a confidence interval size but allowed for different means in the predicted distribution. This table shows the percentage of measured $log(RV)$ which were in the predicted confidence interval for selected returns above a threshold. The selection corresponds to only positive moves. The first columns shows the average return,  mean$( r )$, of those selected points. The confidence intervals are selected to be smaller than twice the average-return of the selected points, here mean$( r )$. This table highlights the ability of our model to predict large upward moves of any size.}
\end{subtable}

\end{table}

\subsection{Evolution of the coefficients}
\begin{figure}[h]
\centering
\includegraphics[width = \figurewidth, height = \figureheight]{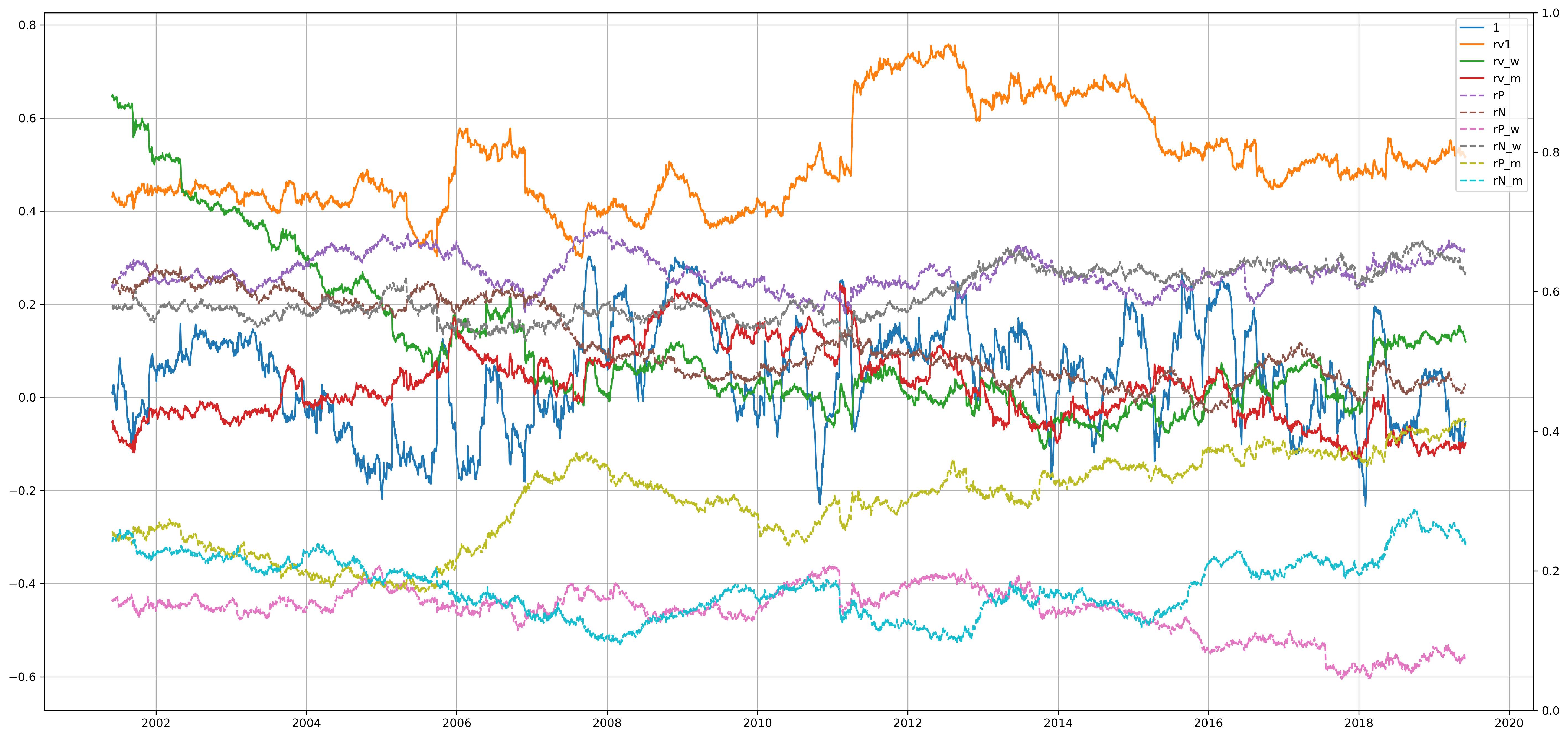}
% \vspace{-10pt}
\caption{\label{fig:IntCoeff} This graphic shows the evolution of the endogenous coefficients for the company ACS. The solid lines correspond to the $RV$ at different frequencies while the dashed ones are for the leverage effect coefficients. At each time $t$ this figure shows which variable is mainly driving the volatility. In addition, the sequential update of these coefficients allows us to observe on this figure how they evolve through time.}
\end{figure}

The algorithm was built to explain what is driving the time series by decomposing it into different economic factors. We assessed the accuracy of this decomposition in the previous section by measuring the quality of the direct one step ahead inference. Going into more details, Figure \ref{fig:IntCoeff} shows the evolution of the endogenous variables through time for the company Actividades de Construccion y Servicios SA (ACS), which we chose randomly out of the $487$ stocks. Hence, at any time $t$ this graphic shows which variable and frequency is mainly influencing the stocks' behaviour. We did not put the corresponding graphic obtained by the HAR-RV model since in that case only the coefficient of the previous RV value was meaningful throughout the dataset.\bigskip
% It is interesting to observe their evolution and the change of behaviour that occurred after the financial crisis. We did not put the corresponding graphic obtained by the HAR-RV model since in that case only the coefficient of the previous RV value was meaningful throughout the dataset.\bigskip

\begin{figure}[h]
\centering
\includegraphics[width = \figurewidth, height = \figureheight]{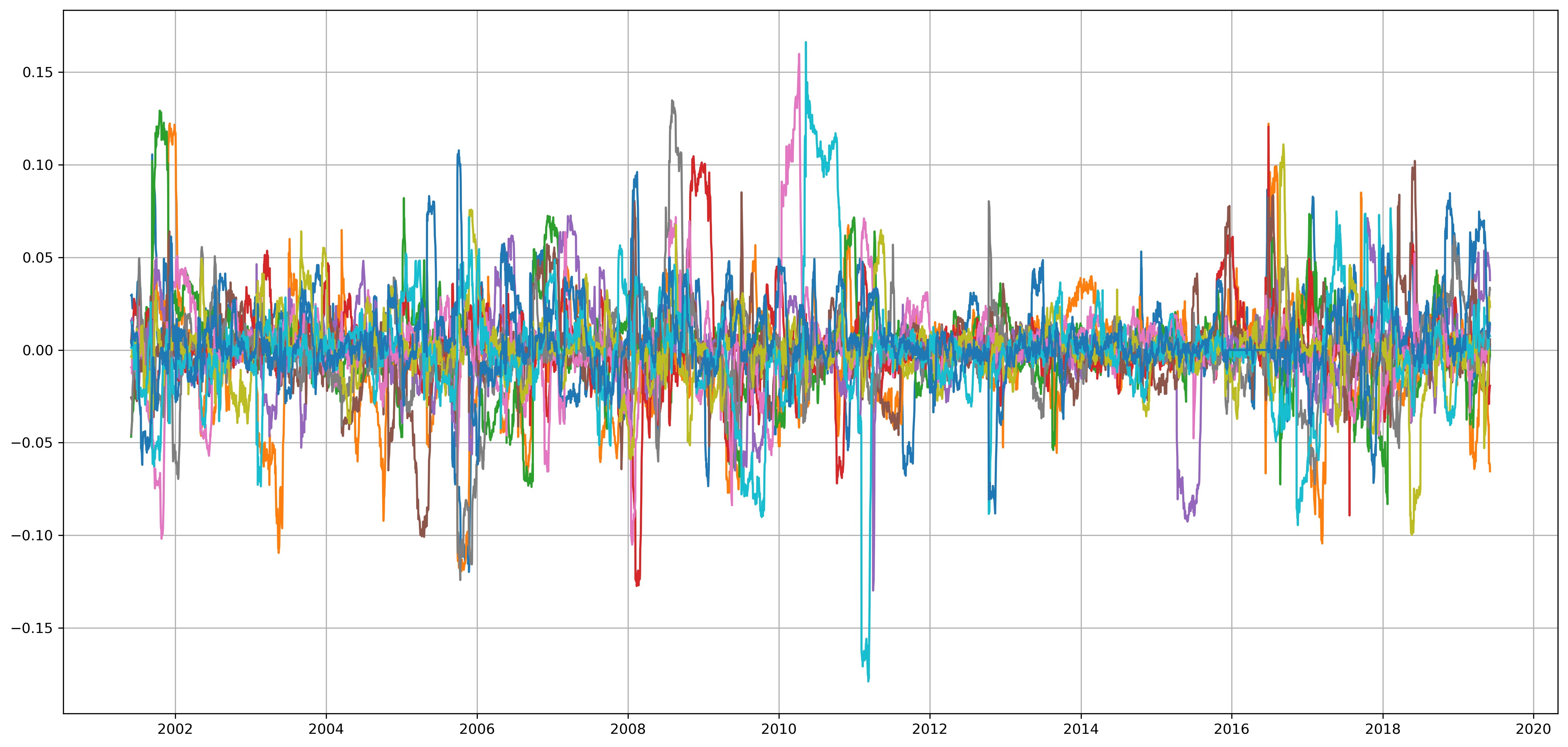}
% \vspace{-10pt}
\caption{\label{fig:ExtCoeff} This figure shows the evolution of the exogenous coefficients for the company ACS. At each time $t$ the lines correspond to the value of the coefficients of the selected core-parent group variables. It shows at each time $t$ which parent is mainly influencing the volatility of ACS. Also, it shows how the coefficients of the different exogenous variables evolve through time.} %; a good example is the bump during the financial crisis of $2008$.}
\end{figure}

Figure \ref{fig:ExtCoeff} shows the evolution of the exogenous coefficients through time for the company ACS. The variables correspond to the parents which are selected dynamically with the two step process described in Section \ref{sec:sec_pa}. Hence, as the coefficients evolve with the selected parents, the new ones will start close to zero. This effect is responsible for the observable mean reversion on the graphic. As with the endogenous variables, at any time $t$ this graphic is showing which stock in the market is mainly influencing the move. An interesting aspect we can see on this Figure \ref{fig:ExtCoeff} is the decrease in activity of these coefficients between $2010$ and $2016$ after the financial crisis. This observation motivated us to study how the mean absolute values of these coefficients evolve through time.\bigskip
% At a lower scale it is interesting to see how the  absolute values of those coefficients evolve.\bigskip

\begin{figure}[h]
\centering
\includegraphics[width = \figurewidth, height = \figureheight]{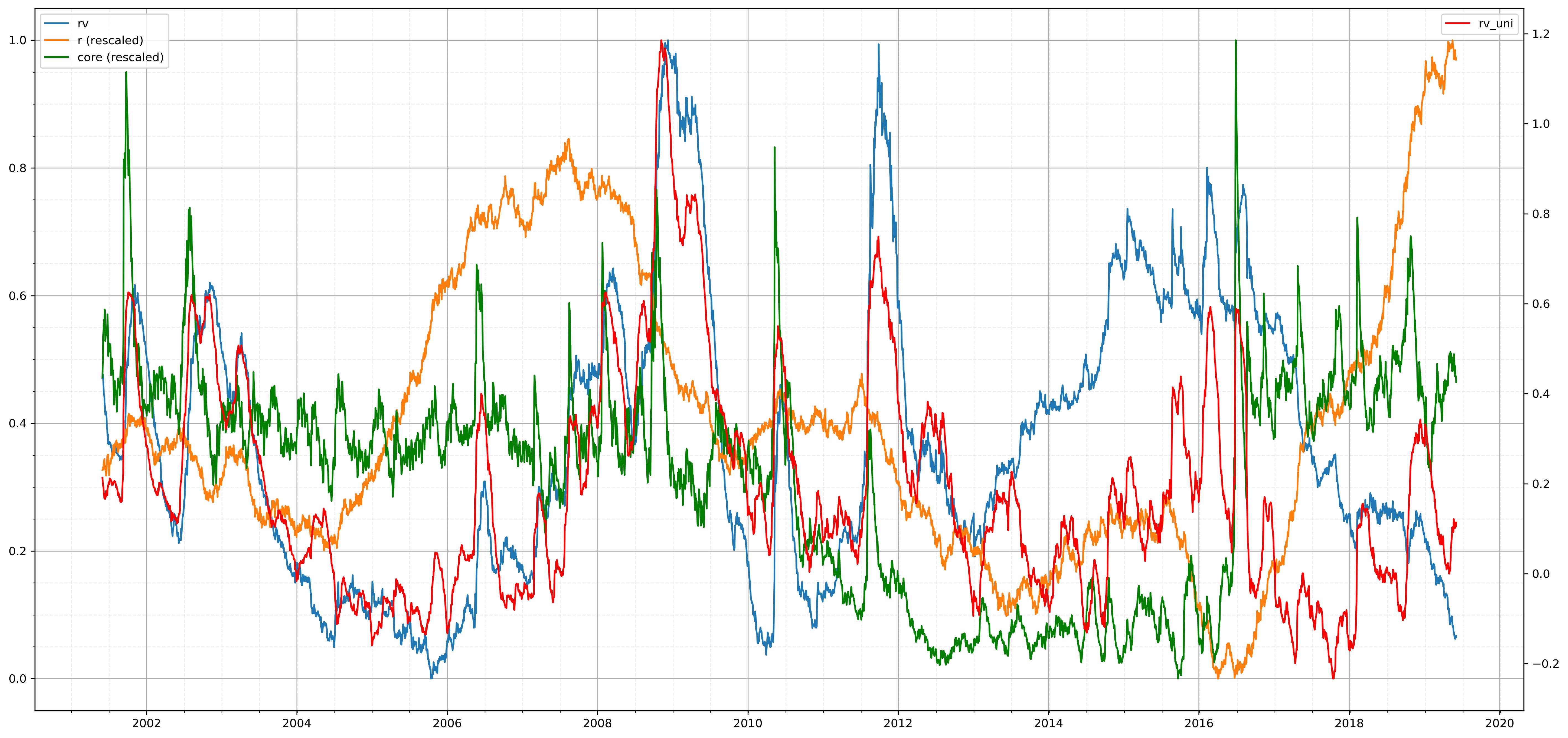}
% \vspace{-10pt}
\caption{\label{fig:FullCoeff_EU}Plots computed with data from $500$ European stocks. The graphic shows the evolution of the different groups (endogenous, with RV and R, and exogenous, denoted Core) rescaled to be on the same $y-axis$, which is left on the figure. The right axis corresponds to the approximated $log(RV)$ of the market, i.e. an equal weighted average of the individual $log(RV)$. This graphic highlights the connections between the behaviour of the market, represented here by $log(RV)$, and the different groups: RV, R and core. For example, looking at the volatility jump of the financial crisis of $2008$ the RV and Core group seem positively correlated with the market while the leverage group seems negatively correlated. What this graphic cannot show is if these connections have any predictive power over the volatility of the market or if they simply react to it with a time lag, Figure \ref{fig:perfAllvsUni_EU} shows this predictive relation.}
\end{figure}

To understand how these patterns are related to the underlying time series we clustered the coefficients into three groups: a first group including all the RV variables, $rv$, a leverage one for the leverage effect coefficients, $r$, and an exogenous group, $core$. These groups are constructed by summing the absolute values of the individual coefficients. We then sum the values of the individual groups to obtain a market view. The log-realised-variance, $log(RV)$, of the market is modelled by an average over the individual ones. Figure \ref{fig:FullCoeff_EU} shows the evolution of the different groups and the market on the same graphic. Even from this low scale view we can observe correlations between the behaviour of these different variables; not only between the market and each group separately, but also between the market and the spread between the different groups.\bigskip

Before commenting on Figure \ref{fig:FullCoeff_EU}, it is interesting to think about what we should observe. We expect a stock to behave differently when its move is driven more by exogenous factors then endogenous ones and vice-versa. More interestingly, when the behaviour of a stock is changing, e.g. going from endogenously to exogenously driven, that should impact its variance. One interpretation is that when the variance of the stock is driven by endogenous factors its behaviour is mainly influenced by internal information and thus less sensitive to the market. Hence, it should be more predictable which should result in a lower variance. On the other hand, when the contribution of the exogenous factors increases that may be correlated with an increasing influence of the market and thus uncertainty on that company which should result in higher variance.\bigskip

In Figure \ref{fig:ExtCoeff} we could observe a recent increase in activity of the exogenous variables from the end of $2016$. In the market view of Figure \ref{fig:FullCoeff_EU} we can clearly see the spike of the exogenous group at the end of $2016$, followed by a spike in the variance of the market. While Figure \ref{fig:FullCoeff_EU} is interesting for having a global view and defining the relations between the different groups and the market, the approach described in the next section will give a clearer view of the strength of this relation.

\subsection{Change point algorithm}
We want to leverage these correlations between the different groups and the underlying time series to make predictions. To do so we will focus on predicting either an increase, decrease or neutral move. Since we previously observed that each group seems to be correlated to the time series at different frequencies we used a simple discrete derivative approximation to detect a change of trend for each of those frequencies. The derivative is approximated by the difference between the value at $t$ and with a lag at $t-l$. We only considered multiples of $5$ days for the lag $l$. For example, for daily moves we used the data up to year $2010$ to find time lag $l$, for which the endogenous group was most correlated to $log(RV)$; this turned out to be one month $l=20$. At each time $t$, we look at the sign of the difference between the value of the endogenous group at time $t$ and $t-20$ to infer whether the $log(RV)$ will increase, decrease or stay neutral for the next day.\bigskip

With this set-up we looked at the correlation between the moves of the groups and $log(RV)$ at different frequencies: daily, two days in advance, weekly, by-weekly and monthly; i.e. lags=$\{1,2,5,10,20 \}$. We used this change point approach to produce inferences on the variance at those different frequencies. Each signal produces a value in the set $\{-1, 0, 1 \}$ which corresponds respectively to a predicted decrease, constant or increase in variance at the frequency under study. Hence when we use a monthly lag, i.e. $20$ we take one points of the data set every $20$ and use the difference of the signals to predict the move for the next time point, i.e. $20$ days later. In other words, we make one month ahead prediction. Interestingly, each group produces their most accurate prediction at different frequencies. Since each signal is evolving at a different frequency and related to a different economic information, they provide different insight on the behaviour of the underlying time series. We sum all of them to obtain the final inference. To reduce the signals' volatility we can add a threshold representing the confidence on the predicted value. For example, by putting a threshold of $2$, the absolute sum of the signals must be higher than $2$ for a position to be entered i.e. more than $2$ different signals need to agree on the same direction.\bigskip

\begin{figure}[h]
\centering
\includegraphics[width = \figurewidth, height = \figureheight]{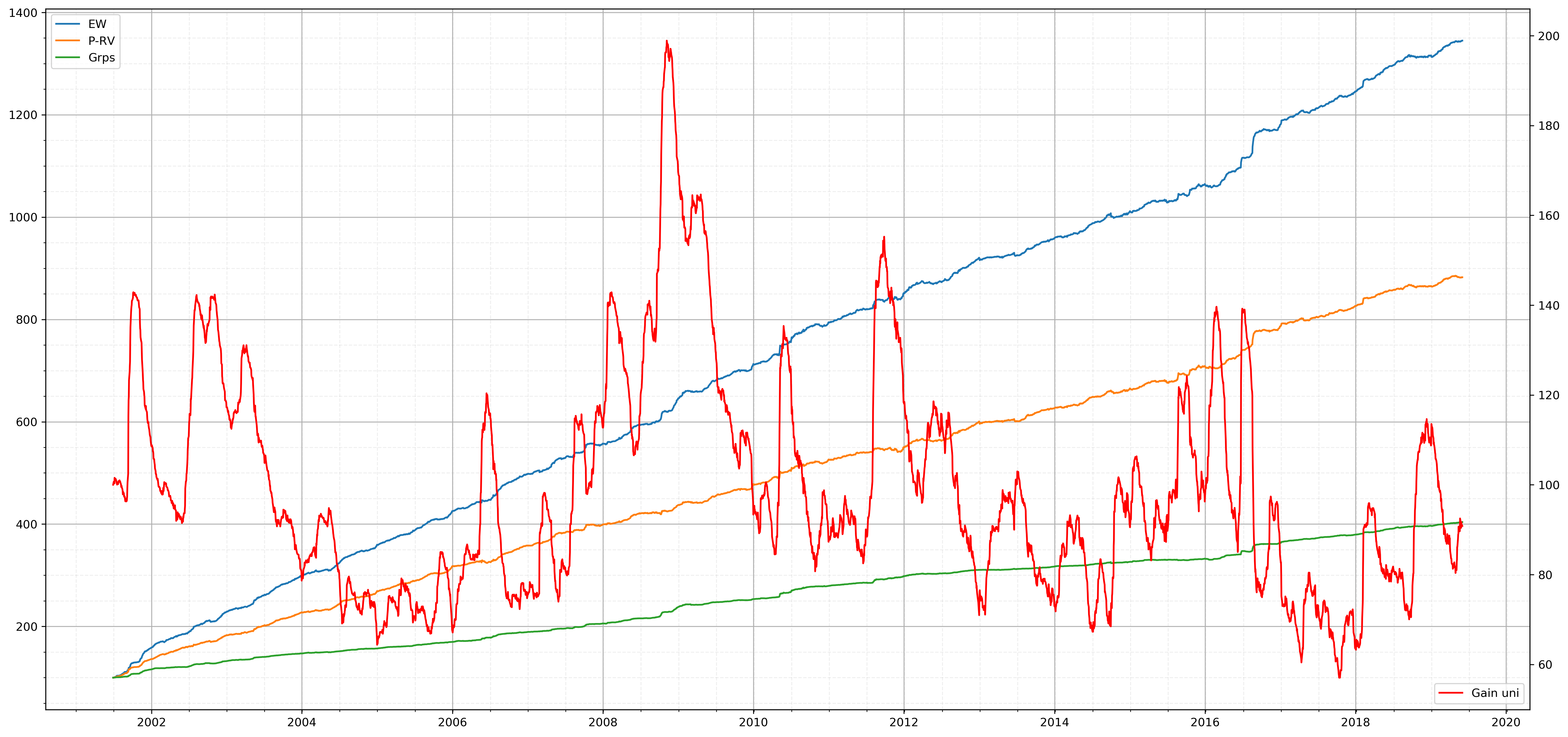}
% \vspace{-10pt}
\caption{\label{fig:perfAllvsUni_EU}Plots computed with data from $500$ European stocks. The orange line shows the performance of an EW portfolio built from the one-day-ahead inference. The green line corresponds to the performance of the EW portfolio using the signals from the change point algorithm applied on the different groups. The blue line shows the performance of the EW portfolio built by combining the one-day-ahead direct inference with the signals from the change points at different frequencies. The simulated portfolios correspond to equal weighted portfolios over the individual time series. In order to highlight the constance through time of the quality of the inference we added in red and on the right Y-axis the evolution of the underlying $log(RV)$ of the market.}
\end{figure}

\begin{figure}[h]
\centering
\includegraphics[width = \figurewidth, height = \figureheight]{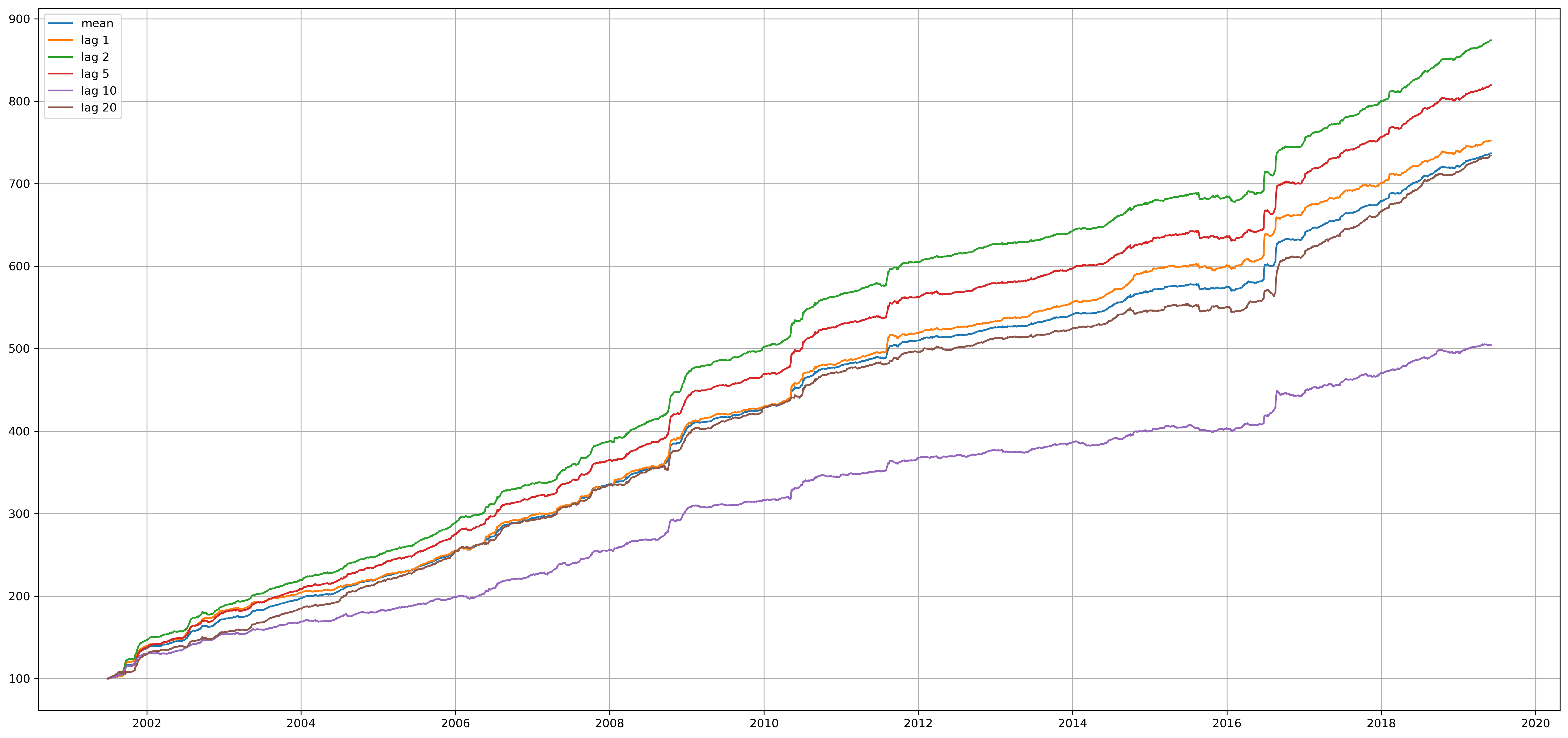}
% \vspace{-10pt}
\caption{\label{fig:perfRVdiffLag_EU}Plots computed with data from $500$ European stocks. This figure plots the performance of the EW portfolio computed with the signals from the group of external coefficients (core group) created with the change points at different frequencies. This graphics shows that while the portfolios using signals updated at lower frequencies than daily are more volatile they still have a linear performance through time and small draw-downs.}
\end{figure}

Using the previously described logic, we performed a backtesting on the underlying $log(RV)$ to demonstrate the quality of the prediction. In order to assess the performance for the whole environment we constructed an equally weighted portfolio and compared it to the market. The linearity observable in Figure \ref{fig:perfAllvsUni_EU} shows that the quality of the prediction is constant through time. Also, if we allow for only two states be it either short-only, $\{-1, 0\}$, or long-only, $\{0, 1\}$, they both perform as well, hence the prediction does not show any bias toward a certain move direction. Another interesting aspect resides in the prediction for different time scales. Running the change point for different time scales on the group of $RV$ coefficients and computing the EW portfolio for each of them produces Figure \ref{fig:perfRVdiffLag_EU}. It represents the performance of the EW portfolio using signals created with the coefficients studied at the different lags (in days) $\{1, 2, 5, 10, 20 \}$. E.g. a lag $s=20$ corresponds to an update of the prediction every $20$ days. The usual issue with long term predictions is the potential size of the drawdowns due to the re-sampling frequency. While more volatile than the lower frequencies the portfolio with monthly update has small draw-downs.\bigskip

\begin{figure}[h]
\centering
\includegraphics[width = \figurewidth, height = \figureheight]{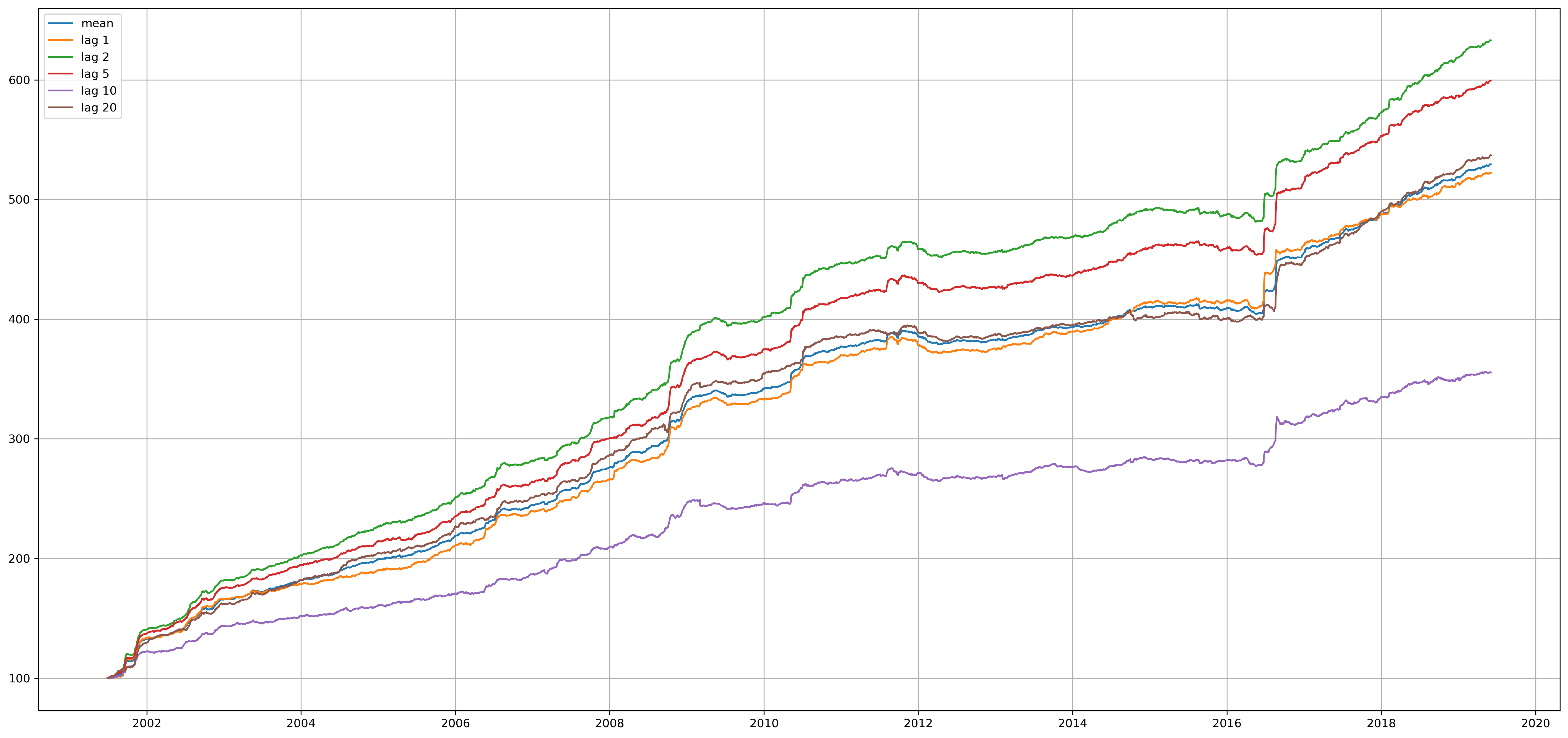}
% \vspace{-10pt}
\caption{\label{fig:perfDiffExtmInt_EU}Plots computed with data from $500$ European stocks. The figure shows the performance of the EW portfolio computed with the signals from the difference between the external and the RV groups using the change points algorithm at different frequencies. Hence those plots show the positive correlation between the difference  \textit{(exogenous coefficients - endogenous  coefficients)} variables and the evolution of the volatility. In other words, when the volatility becomes more driven by exogenous factors than endogenous ones the volatility increases and vice-versa.}
\end{figure}

Let us recall the question raised in the previous section: is the variance increasing when the spread between the exogenous and endogenous variables increases or decreases? Following this change point set-up Figure \ref{fig:perfDiffExtmInt_EU} shows the difference between the exogenous group minus the endogenous ones for the European stocks we are studying. For the different frequencies we looked at, $\{1, 2, 5, 10, 20 \}$, the variance increases when the exogenous groups become more important than the endogenous one, and vice versa. In other word the behaviour of the variances is positively correlated with the spread exogenous minus endogenous groups. This corresponds to the results we expected, i.e. the more a variance is being driven by external factor the more uncertain its behaviour is and thus its variance increases, while if the endogenous coefficient are dominants the stock moves due to information proper to itself and thus in a more predictable manner which decreases the variance. However, as we can see on Figure \ref{fig:perfDiffExtmInt_EU} the quality of that connection is not constant with time and depends on the market conditions.

\subsection{Risk metric application}
\label{sec:results_risk}
\begin{figure}[h]
\centering
\includegraphics[width = \figurewidth, height = \figureheight]{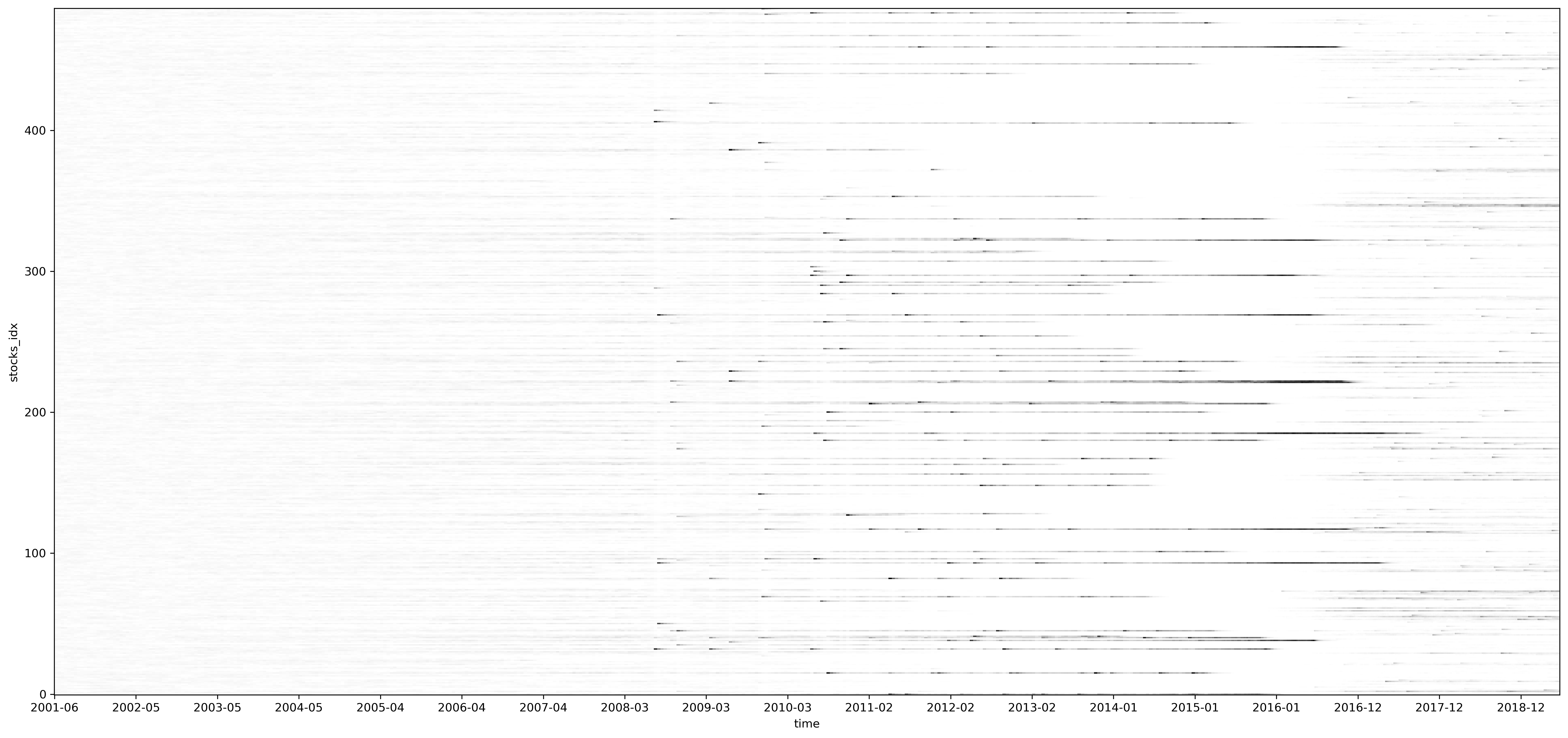}
% \vspace{-10pt}
\caption{\label{fig:paMat_EU}Plots computed with data from $500$ European stocks. The $y-axis$ corresponds to the index of each of the $487$ stocks while the time is on the $x-axis$. This graphic is a $2$-dimensions representation of the evolution of the matrix of core-parental-set through time. For each stock, at each time $t$ the darker the point the more core-parent sets it is part of. Hence, with this graphic we can observe at each time $t$ which stocks are influencing many others, that corresponds to the ones with dark points. While, the stocks in white are followers; i.e. they are not influencing the move of any other stock in this environment.}
\end{figure}

While the previous section focused on the direct inference of $log(RV)$, this section will discuss the application on market stress. The SGDLM model uses a multivariate Wishart distribution to compute a graph of the market and select the parents for each DLM. With this selection each DLM only has a few parents representing the stocks with the highest influence. We expect to see stocks that are driving a sector and others that are followers. In Figure \ref{fig:paMat_EU} we plotted the evolution of the matrix of parents' relation through time. At each time $t$ we counted the number of core parent sets each stock is part of. The higher the number of sets the blacker and vice-versa. By doing so, we want to identify the stocks that are in many core-sets and thus influencing the environment. Indeed, by looking at this matrix we can identify few stocks that have a clear influence in the market. Another way to see this would be to consider the market risk to be concentrated on a few stocks which moves could have systemic impacts. We can also observe a clear change occurring after the financial crisis and after $2016$.\bigskip

The Figure \ref{fig:paMat_EU} is an interesting argument in favour of an asymmetric sparse covariance matrix. Indeed with the classic multivariate approach modelling the cross-relationships with a symmetric covariance this matrix would not exist. This assymetry exists because the SGDLM decomposes the parent update into two phases, the first using the multivariate Wishart to select the potential candidates from the precision matrix; and the second computing the variables' coefficients as a regression in each individual DLM to predict the move of the stock and selecting the parents according to their signal-to-noise ration. And for a stock to stay in the core group, which Figure \ref{fig:paMat_EU} represents, it needs to pass the two phases i.e. have a high precision element in the multivariate Wishart, and a high signal-to-noise ratio when included in the up set of the SGDLM and part of the DLM's variables. Thus, the coefficients of the core-group variables represent the strength of the relationship between the stock and its selected parents; and the parents selected in the core group the stocks with the highest signal-to-noise ratio.\bigskip

In \cite{Corsi:2008ei} they modified the original HAR-RV to model the vol-of-vol, or variance of the variance. This vol-of-vol represents the variance of the DLM and hence the confidence in its prediction and decomposition. If that variance increases, either it is due to a bad fit from the DLM or it represents an increase in uncertainty for the variance of that stock. The inferred variance of variance is on average much lower than the computed one although it does spike during jumps of variances. In addition, the percentage of correct predictions shown in table \ref{tab:confInt_EU} when restricting the analysis to large moves is another measure of the quality of the inferred variance of variance.\bigskip

\begin{figure}[h]
\centering
\includegraphics[width = \figurewidth, height = \figureheight]{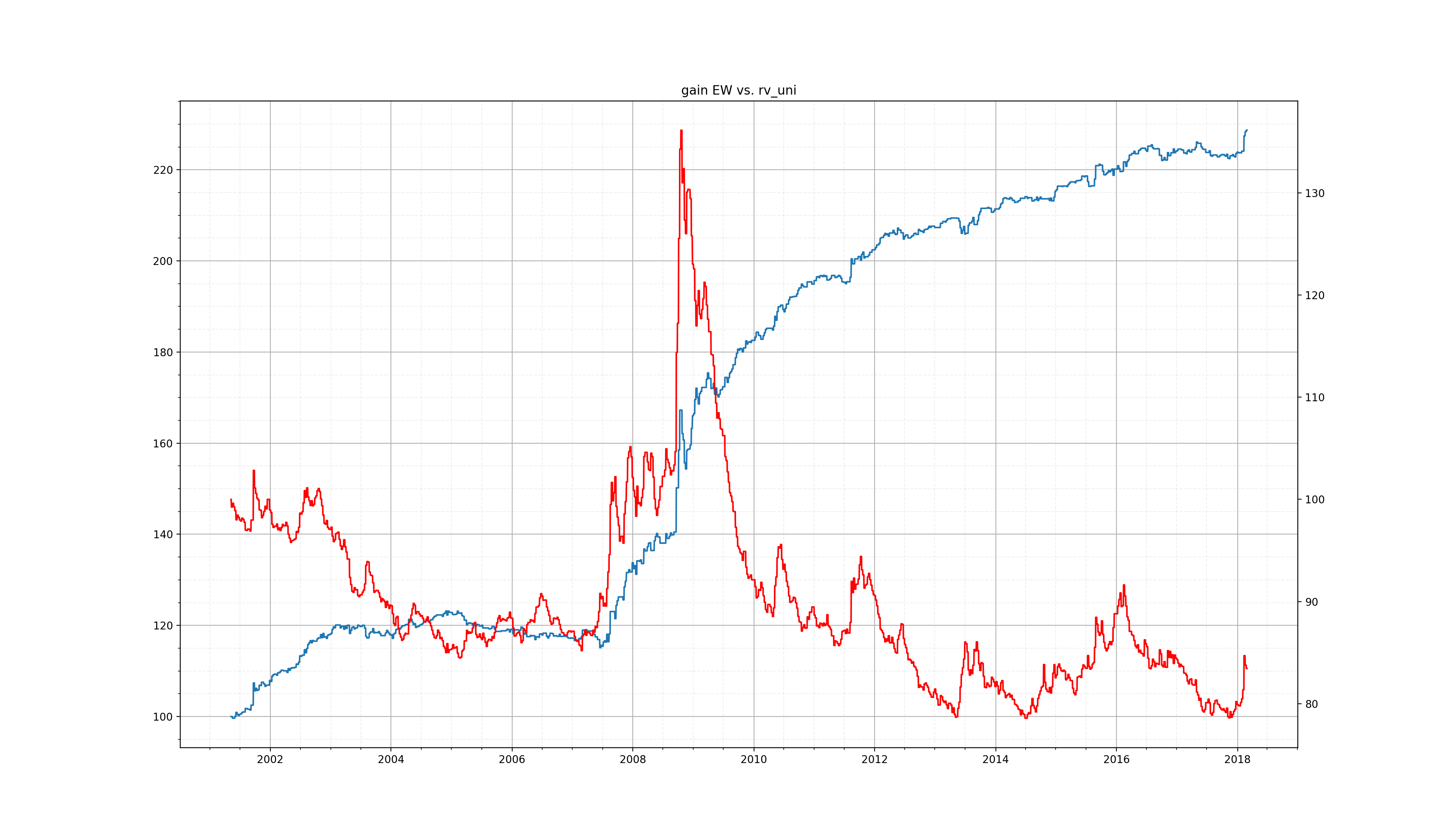}
\vspace{-20pt}
\caption{\label{fig:perfStlfsiEWth}Plots computed with $390$ stocks from the S\&P$500$. The left axis is linked to the blue line that represents the performance of the portfolio computed with the signals from the change point algorithms using the evolution of the different groups to predict the evolution of the STLFSI index. In red and linked to the right axis is the STLFSI stress index re-based to $100$ at the start of the backtest. This graphic shows the quality of the prediction of the stress index using the signals obtained from the different groups from $390$ stocks from the S\&P$500$.}
\end{figure}

\cite{Gruber:2016wn} used the KL divergence as a market stress indicator and compared it to the St. Louis Fed Financial Stress Index (STLFSI). They observed that both metrics increased at the same time to the point of considering their metric to be better than the STLFSI index since it reacted faster. It turns out this index is quite similar to the computed estimate of the market $log(RV)$. Thus, we used the change point algorithm on the different group variables described previously to produce an inference of this index. As explained previously, we can add a threshold on the value of the sum of the added signal, to show how that can impact the EW portfolio; Figure \ref{fig:perfStlfsiEWth} used a threshold of $4$, which means the absolute sum of the signals must be higher than $4$ for a position to be taken. Enforcing such a high threshold diminishes the number of trades. But, Figure \ref{fig:perfStlfsiEWth} shows the quality of the prediction after $2007$. It is not clear why the prediction between $2004$ and $2007$ stagnate, could be due to the low variability of the STLFSI itself.\bigskip

%%%%%%%%%%%%%%%%%%%%%%%%%%%%%%%%%%%%%%%%%%%%%%%%%%%%%%%%%%%%%%%%%%%%%%%%%%%%%%%%%%%%
\section{Application of the algorithm on different environments}
\label{sec:resultsDiffEnv}
\subsection{Application on the stocks from the S\&P$500$}
\label{sec:resultsEU}
With exactly the same set-up as in section \ref{sec:results} for EU stocks, i.e. without OHLC data, we ran the algorithm on US stocks from the S\&P$500$. While we used different stock exchanges and countries to select the most liquid European stocks for the US market we use a different methodology. We select the stocks currently present in the S\&P$500$ index but only select the $376$ stocks that have a history since $2000$.\bigskip
% Since not all these stock have a price history going back to the year $2000$ we assumed $log(RV)$ to be zero when the price is missing. Thus, the number of stocks increases from $368$ in $2000$ to $500$ after $2018$. We chose to have an evolving size of dataset to see how the algorithm adapts to it. As for the European stocks, with that number of time series the available GPU memory could only do $500$ Monte Carlo samples.\bigskip
% The data corresponds to the $497$ most liquid stocks selected from thousands of stocks in different European markets. We start the backtest from $2000$ but not all stocks had such a long history. Thus, contrary to what we did for the stocks from the S\&P$500$ we kept all of them and assumed $log(RV)$ to be zero when the price is missing. With that number of series the available GPU memory could only do $500$ Monte Carlo samples.\bigskip

Because all the stocks in this environment belong to the same stock index we expect to see more connections between the time series in the index and thus the SGDLM model to perform much better than in the European environment which mixed many stock exchanges and countries. In terms of metrics the direct one step ahead prediction does not perform as well as for the European environment which is probably due to the lower amount of information contained in the HAR variables. We followed the same methodology as described in section \ref{sec:results_met_EU} for European stocks. As for the European stocks, in the case of large moves our model correctly predicted more than $60.0\%$ of the moves, positive or negative. While it correctly predicted $65.93\%$ of moves bigger than $11.25\%$ and $63.83\%$ of the increases in variance larger than $20.27\%$. On the same thresholded-data, the classic HAR-RV never achieves more than $55.86\%$ accuracy and, while the SGDLM performance is close to our  model for small moves the gap increases proportionally to the size of the moves considered. Table \ref{tab:confIntabs_US} shows the percentage metric for moves with an absolute return larger than a certain threshold. And, Table \ref{tab:confInt_US} the percentage of correct predictions when focusing on volatility increases alone.\bigskip

\begin{table}[h]
\centering
\caption{Performance metrics of the one-day-ahead inference obtained for stocks from the S\&P$500$.}

\begin{subtable}{\textwidth}
\centering
\begin{tabular}{ l | c | c | c | c}
\multicolumn{1}{c | }{Model} & t - 1 & HAR-RV & SGDLM & Full \\
\hline \hline
Median ADV         & $0.011$  & $0.014$  & $0.018$  &  $0.015$\\ 
Median RMSE        & $0.040$  & $0.041$  & $0.047$  &  $0.044$\\
Median $R^2$       & $0.991$  & $0.991$  & $0.988$  &  $0.990$\\
 M-Z coefficient  & N/A       & $0.995$  & $1.007$  &  $0.996$\\
\end{tabular}
\caption{\label{tab:mad_US} One day ahead inference comparison between different models. Median ADV corresponds to the Median Absolute deviation. Median RMSE is the median of the Root Mean Squared Errors. M-Z coefficient corresponds to the Minor-Zarnowitz regression coefficient, i.e. the regression of the observed values on the predicted ones.}
\end{subtable}

%\bigskip
\end{table}

\begin{table}[p]
\centering
\caption{Performance metrics with confidence interval of the one-day-ahead inference obtained for $390$ stocks from the S\&P$500$.}

\begin{subtable}{\textwidth}
\centering
\begin{tabular}{ l | l | c | c | c }
\multicolumn{2}{c | }{Model} & HAR-RV & SGDLM & Full \\
\hline \hline
\multirow{2}{*}{without threshold}  & Interval size   & $4.37\%$ & $4.37\%$ & $4.39\%$ \\ 
  								&  \% in  & $59.02\%$ & $62.47\%$ & $62.57\%$ \\
\hline
\multirow{2}{*}{with  mean$(| r |) = 5.74\%$ } &  Interval size  & $10.18\%$ & $10.16\%$ & $10.19\%$ \\
  									      &  \% in   & $53.72\%$ & $60.32\%$ &  $61.06\%$ \\
\hline
\multirow{2}{*}{with mean$(| r |) = 11.25\%$ }  &  Interval size  & $22.07\%$ & $22.00\%$ &  $22.04\%$ \\
  									     &  \% in   & $54.21\%$ & $64.56\%$ &  $65.93\%$ \\
\hline
\multirow{2}{*}{with mean$(| r |) = 19.96\%$ }  &  Interval size  & $39.35\%$ & $39.34$ &  $39.40$ \\
  									      &  \% in   & $50.23\%$ & $61.00\%$ &  $63.26\%$ \\
\end{tabular}
\caption{\label{tab:confIntabs_US}In order to compare the different models, we fixed a confidence interval size but allowed for different means in the predicted distribution. This table shows the percentage of measured $log(RV)$ which were in the predicted confidence interval for the selected absolute-returns above a certain threshold. The first columns shows the average absolute-return,  mean$(| r |)$, of those selected points. The confidence intervals are selected to be smaller than twice the average-return of the selected points, here mean$(| r |)$. This table shows the ability of our model to predict large, positive and negative, moves of any size.}
\end{subtable}

\bigskip

\begin{subtable}{\textwidth}
\centering
\begin{tabular}{ l | l | c | c | c }
\multicolumn{2}{c | }{Model} & HAR-RV & SGDLM & Full \\
\hline \hline
\multirow{2}{*}{with  mean$ (r ) = 6.97\%$ } &  Interval size  & $12.18\%$ & $12.18\%8$ & $12.19\%$ \\
  									    &  \% in   & $55.86\%$ & $59.80\%$ &  $60.49\%$ \\
\hline
\multirow{2}{*}{with mean$ (r ) = 11.92\%$ }  &  Interval size  & $22.02\%$ & $22.05\%$ &  $22.06\%$ \\
  									    &  \% in   & $52.07\%$ & $61.08\%$ &  $62.40\%$ \\
\hline
\multirow{2}{*}{with mean$ (r)  = 20.27\%$ }  &  Interval size  & $40.02\%$ & $40.08\%$ &  $40.08\%$ \\
 									    &  \% in   & $51.13\%$ & $61.96\%$ &  $63.83\%$ \\
\end{tabular}
\caption{\label{tab:confInt_US}In order to compare the different models, we fixed a confidence interval size but allowed for different means in the predicted distribution. This table shows the percentage of measured $log(RV)$ which were in the predicted confidence interval for selected returns above a threshold. The selection corresponds to only positive moves. The first columns shows the average return,  mean$( r )$, of those selected points. The confidence intervals are selected to be smaller than twice the average-return of the selected points, here mean$( r )$. This table highlights the ability of our model to predict large upward moves of any size.}
\end{subtable}

\end{table}

\begin{figure}[h]
\centering
\includegraphics[width = \figurewidth, height = \figureheight]{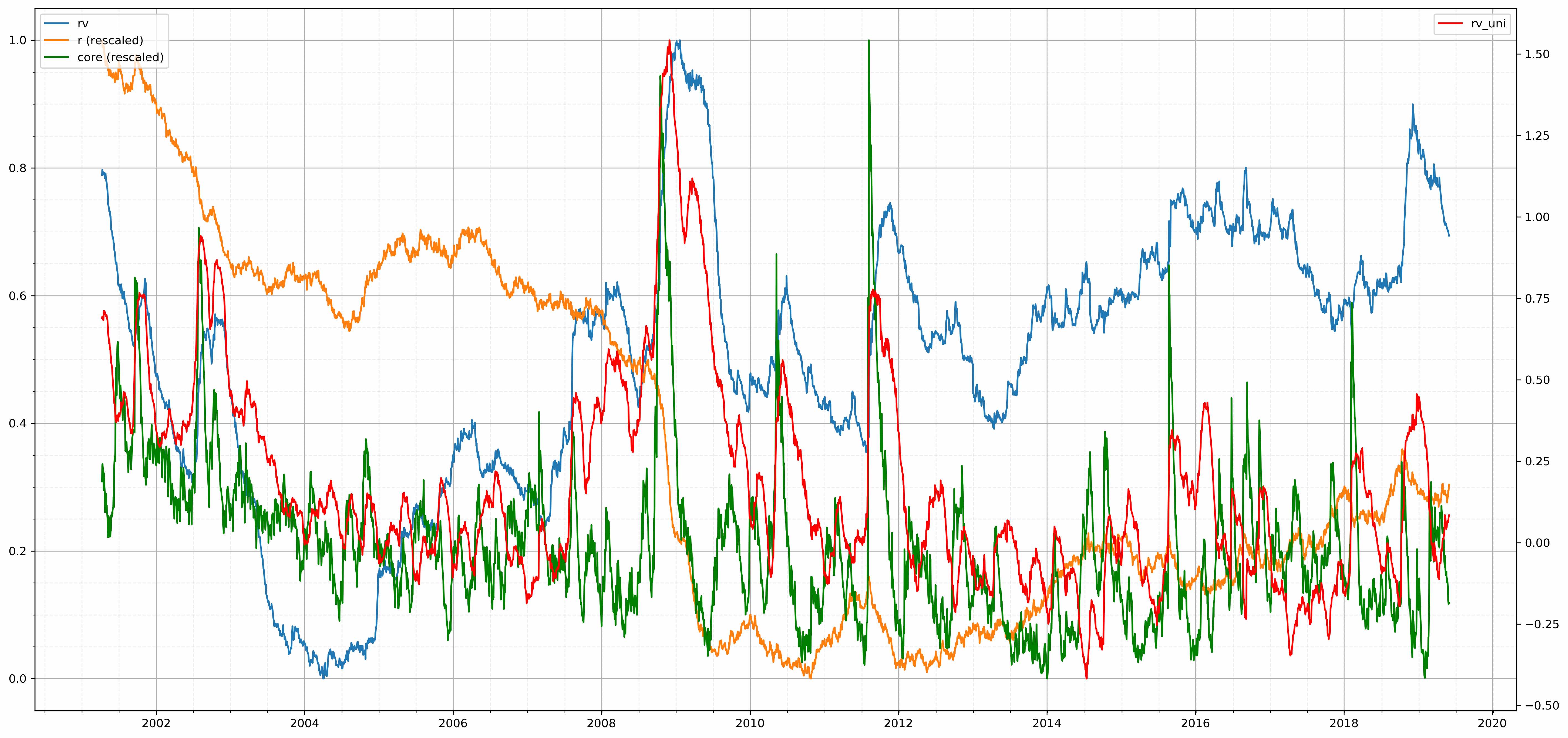}
% \vspace{-10pt}
\caption{\label{fig:FullCoeff_US}Plots computed with stocks from the S\&P$500$. The graphic shows the evolution of the different groups (endogenous, with RV and R, and exogenous, denoted Core) rescaled to be on the same $y-axis$, which is left on the figure. The right axis corresponds to the approximated $log(RV)$ of the market, i.e. an equal weighted average of the individual $log(RV)$. This graphic highlights the connections between the behaviour of the market, represented here by $log(RV)$, and the different groups: RV, R and core. For example, looking at the volatility jump of the financial crisis of $2008$ the RV and Core group seem positively correlated with the market while the leverage group seems negatively correlated. What this graphic cannot show is if these connections have any predictive power over the volatility of the market or if they simply react to it with a time lag, Figure \ref{fig:perfAllvsUni_US} shows this predictive capacity.}
\end{figure}

\begin{figure}[h]
\centering
\includegraphics[width = \figurewidth, height = \figureheight]{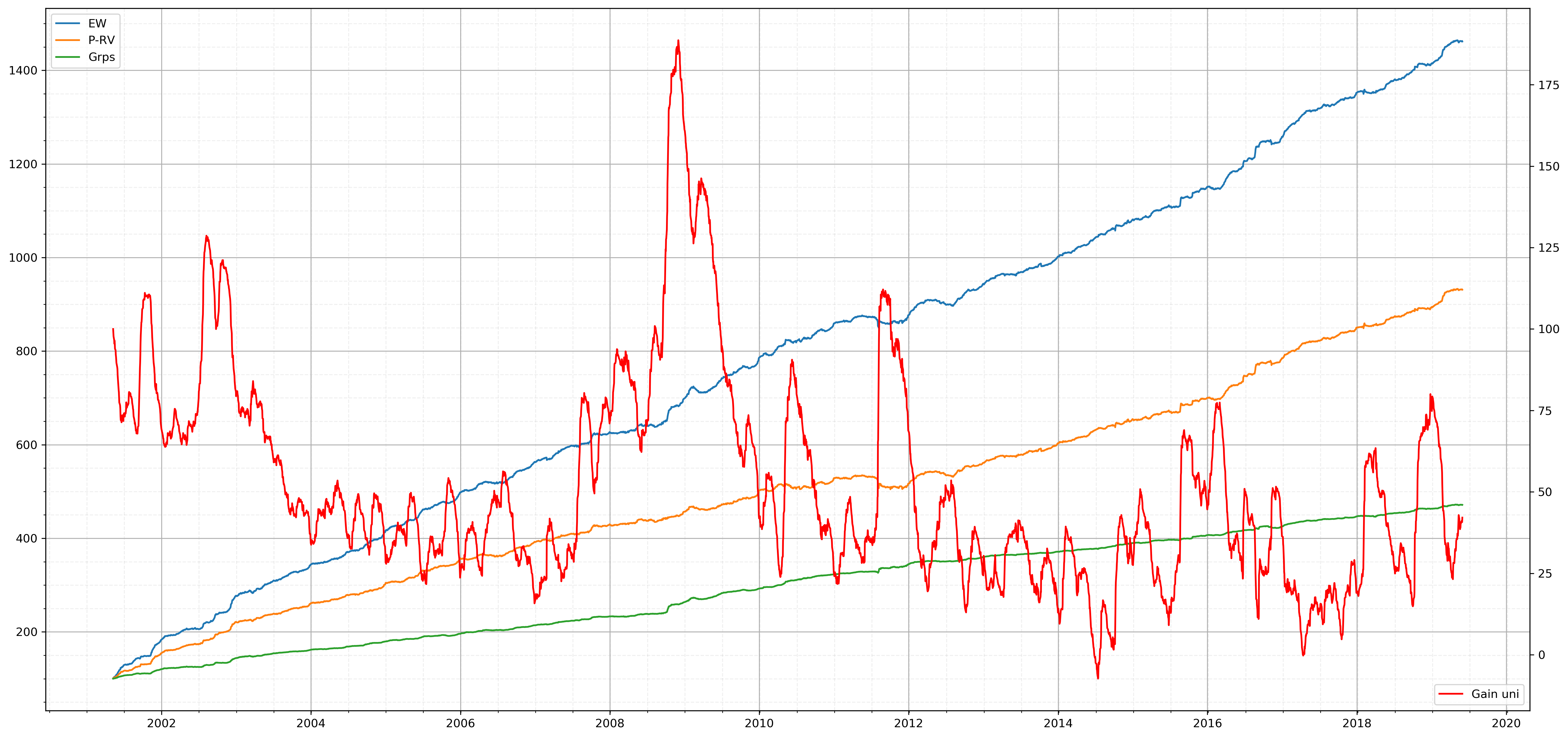}
% \vspace{-10pt}
\caption{\label{fig:perfAllvsUni_US}Plots computed with stocks from the S\&P$500$. The orange line shows the performance of an EW portfolio built from the one-day-ahead inference. The green line corresponds to the performance of the EW portfolio using the signals from the change point algorithm applied on the different groups. The blue line shows the performance of the EW portfolio built by combining the one-day-ahead direct inference with the signals from the change points at different frequencies. The simulated portfolios correspond to equal weighted portfolios over the individual time series. To highlight the consistency through time of the quality of the inference we added in red and on the right Y-axis the evolution of the underlying $log(RV)$ of the market.}
\end{figure}

Figure \ref{fig:FullCoeff_US} shows the evolution of the different groups rescaled to be on the same graph and compared to the averaged $log(RV)$ of the S\&P$500$. Recall that on EU data as shown in Figure \ref{fig:FullCoeff_EU} the evolution of the different groups seemed to be positively correlated to the variance of the environment. On the US data this correlation seems just as strong as we can see in Figure \ref{fig:FullCoeff_US}. We also perform the same backtesting strategy using the signals from the change point algorithm and show in Figure \ref{fig:perfAllvsUni_US} the EW portfolio versus the $log(RV)$ of the market. As previously the performance is constant through time even during the financial crisis, highlighting further the robustness of our model.\bigskip

\begin{figure}[h]
\centering
\includegraphics[width = \figurewidth, height = \figureheight]{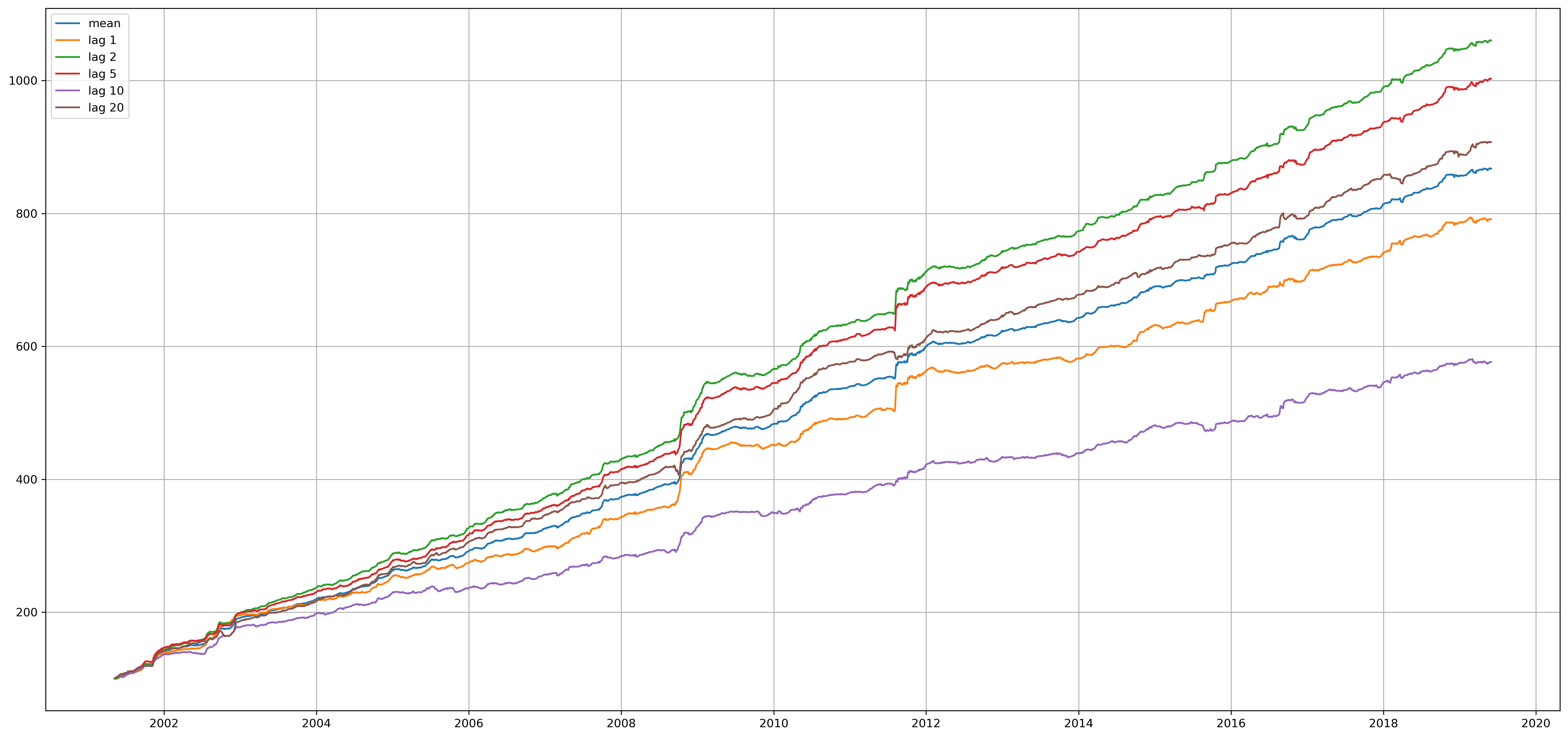}
% \vspace{-10pt}
\caption{\label{fig:perfDiffExtmInt_US}Plots computed with stocks from the S\&P$500$. The figure shows the performance of the EW portfolio computed with the signals from the difference between the external and the RV groups using the change points algorithm at different frequencies. Hence those plots show the positive correlation between the difference  \textit{(exogenous coefficients - endogenous  coefficients)} variables and the evolution of the volatility. In other words, when the volatility becomes more driven by exogenous factors than endogenous ones the volatility increases and vice-versa.}
\end{figure}

As previously, we wanted to answer the question: is the variance increasing when the spread between the exogenous and endogenous variables increases or decreases? To do so we computed the EW portfolio from the signal build for the different frequencies $\{1,2,5,10,20\}$ (in days) using the difference between the exogenous and endogenous groups. Figure \ref{fig:perfDiffExtmInt_US} shows the results for those different frequencies. As in the EU market the variance is positively correlated to the difference between these two groups. And the link seems even stronger than the EU market since the backtest is more linear and smoother than in Figure  \ref{fig:perfDiffExtmInt_EU}. Naturally, on the lower monthly-frequency the connection is not as robust as for the higher ones.

\subsection{Applying the model to predict prices of futures}
\label{sec:results_price}
In the previous sections we focused predicting the log-realised-variance of US and EU stocks without using OHLC data, in this section we will show the results on different environments of futures with OHLC data. It is important to study the scale of the different input-variables when applying this algorithm to different datasets. To guarantee the numerical-stability of the computation the variables should have similar scales. Also, it is important to modify the number of allowed parents to retain sparsity relative to the size of the new environment.\bigskip

While we previously worked with the variance, we can adapt the framework to work directly with prices. Following the logic of the information cascade we replace the different realised variance variables by the mean of the prices at these same different frequencies: daily, weekly and monthly. In addition, instead of incorporating the previous squared returns, we use an exponentially weighted average of the returns over the same frequencies. On the other hand, the leverage effect coefficients are not modified. Thus we keep the decomposition into groups with different and complementary economic information.\bigskip

\begin{figure}[h]
\centering
\includegraphics[width = \figurewidth, height = \figureheight]{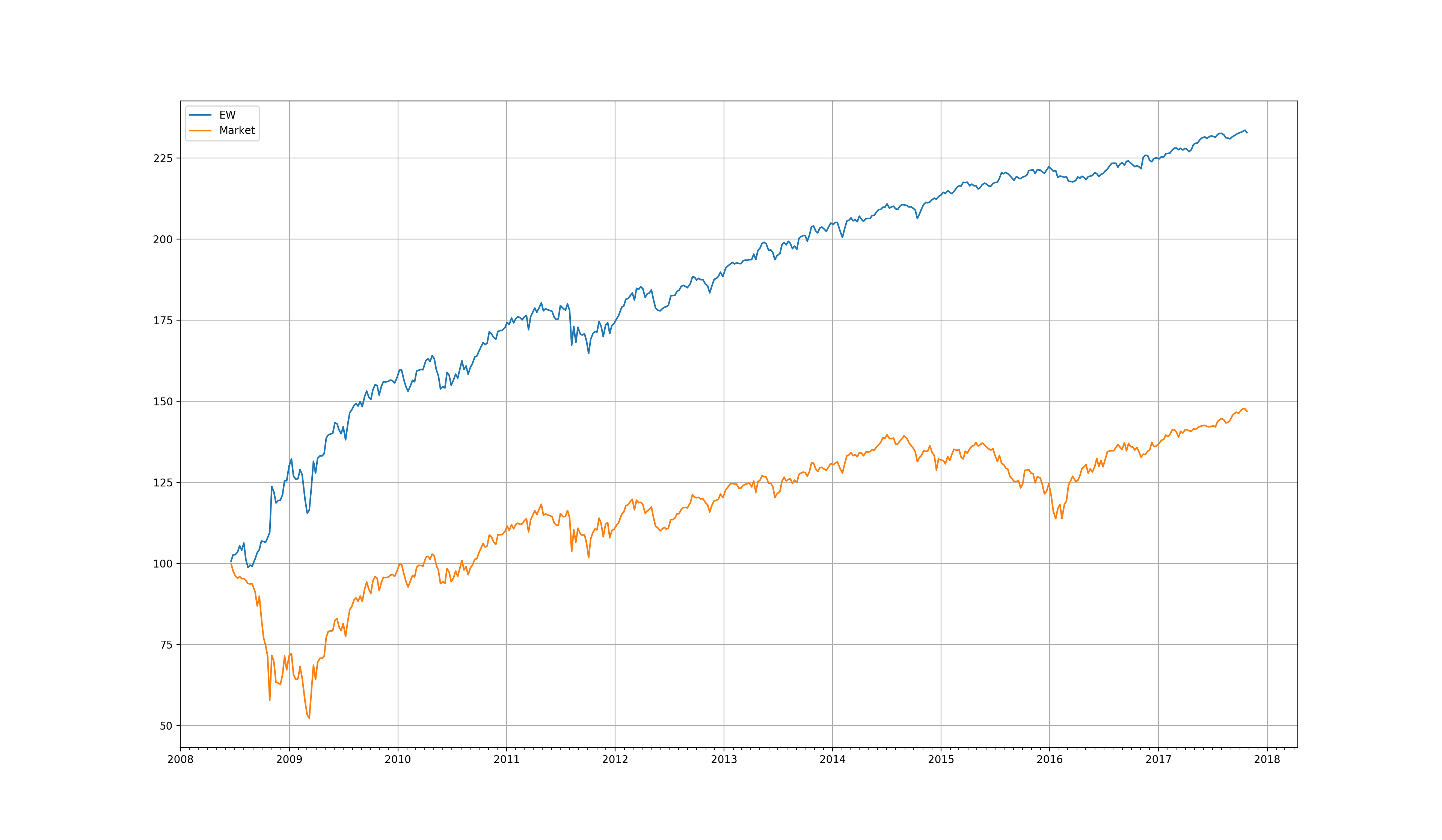}
\vspace{-20pt}
\caption{\label{fig:perfETF}This graphics used data from $75$ futures on ETFs. The yellow line corresponds to the performance of an EW portfolio on all the ETF of the environment. The Blue line corresponds to the performance of our EW portfolio computed using the signals from the change point algorithm on the different groups at different frequencies. The performance of our portfolio shows the quality of the inference especially during large decreases such as the one during the $2008$ crisis.}
\end{figure}

\begin{figure}[h]
\centering
\includegraphics[width = \figurewidth, height = \figureheight]{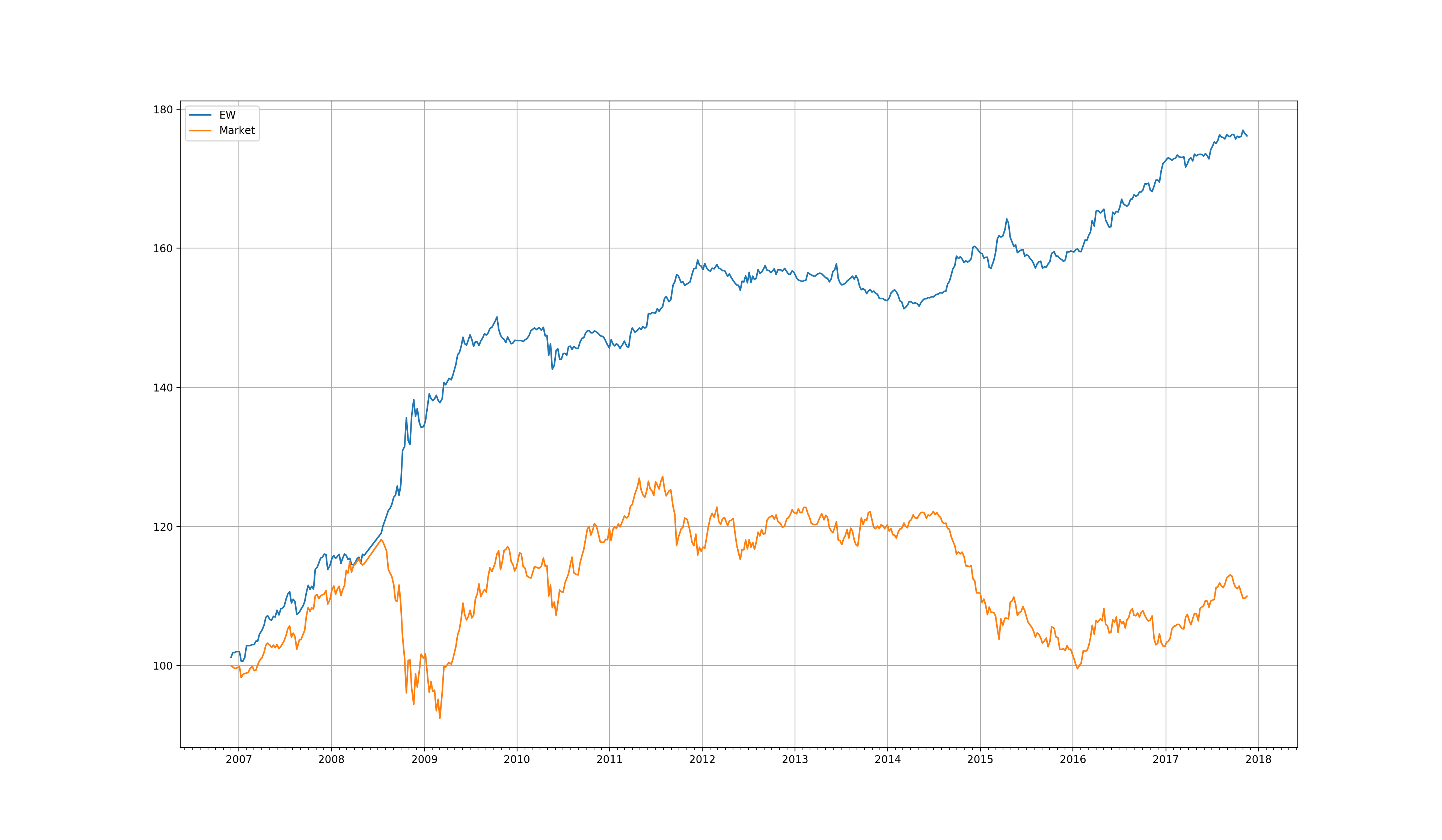}
\vspace{-20pt}
\caption{\label{fig:perfFX}This graphics used data from $22$ futures on ETFs. The yellow line corresponds to the performance of an EW portfolio on all the ETF of the environment. The Blue line corresponds to the performance of our EW portfolio computed using the signals from the change point algorithm on the different groups at different frequencies. The performance of our portfolio shows the quality of the inference especially during large decreases such as the one during the $2008$ crisis.}
\end{figure}

In order to demonstrate the robustness of our approach we now apply the model on ETFs and FX futures. We dispose of $75$ futures on ETF and $22$ on FX-rates. Compared to the previous examples we only had to modify the number of allowed parents to adapt to the smaller dataset. We chose a number of parents in the core set of $5$ for both the ETFs  and FXs environment, but a up-set size of $5$ for the ETFs and $3$ for the FXs. ETF and FX futures have different behaviours from stocks but still follow this influence of endogenous and exogenous information. Hence the model is still able to capture those different connections and create meaningful signals. To show the flexibility and robustness of our proposed model we ran the algorithm on weekly data instead of the daily one. I.e. we selected one point every five days and considered this new dataset as the input for the algorithm. Thus the one step ahead inference now corresponds to one week ahead. We used the EW portfolio to assess the performance on these datasets. The combination of signals built from endogenous and exogenous groups at different frequencies produces a portfolio with low volatility especially on large downward moves, as can be seen in Figures \ref{fig:perfETF} for the ETF and \ref{fig:perfFX} for the FX. As we used weekly data for this example the median length of trades was $2$ weeks for both environments.\bigskip

%%%%%%%%%%%%%%%%%%%%%%%%%%%%%%%%%%%%%%%%%%%%%%%%%%%%%%%%%%%%%%%%%%%%%%%%%%%%%%%%%%%%
% summary of problematic, our solution, our results, future work
\section{Conclusion}
\label{sec:NextSteps}
In this paper we proposed a new model that gives a better understanding of the underlying cause of price fluctuations and what it reveals about a specific asset and the market as a whole.  In particular, we proposed a heterogeneous autoregressive time varying simultaneous graphical multivariate dynamic linear model, H-SGDLM. This model can efficiently model large-scale environments including the cross-series relationships.  It is easy to fit, does not assume stationarity of the underlying time series, and updates itself sequentially. The proposed approach decomposes the price movement into different known underlying economically-meaningful variables to produce a better understanding of the behaviour of the collection of time series; it can explain at any time $t$ which ones are likely to be driving the moves, for example, at which trading frequency and if changes are due to endogenous or exogenous factors.\bigskip

We assessed the quality of our decomposition by looking at the forecasting performance of our model.  We showed that it outperformed previous models in predicting large moves; positive and negative ones alike.  In addition, the quality of these predictions are relatively constant through time, within different markets and for multiple assets types. This decomposition creates new insights into the behaviour of a specific asset and the rest of the market. In order to leverage those new signals, we clustered them into endogenous and exogenous groups. By studying the link between those groups and the underlying time series we could obtain accurate long-term predictions that worked well both for individual assets as well as the market as a whole. This allowed us to produce more than a month ahead inference regarding the evolution of the variance of the asset and the market. Furthermore, these signals were shown to be reliable enough to predict the STLFSI stress index while having learned only on a few hundred stocks.\bigskip

In this paper we focused on using the signals from the endogenous and exogenous groups, although our multivariate model can also produce many other interesting outputs such as an simultaneous directed cross-series graph, illustrating relationships between assets.   This provides many interesting directions for future research using our proposed model. 

%%%%%%%%%%%%%%%%%%%%%%%%%%%%%%%%%%%%%%%%%%%%%%%%%%%%%%%%%%%%%%%%%%%%%%%%%%%%%%%%%%%%
\bigskip
\begin{center}
{\large\bf SUPPLEMENTARY MATERIAL}
\end{center}
\section{SGDLM equations}
\subsection{Posterior and Recoupling}
Each posterior follow a Normal-Gamma distribution $p(\theta_{j,t},\lambda_{j,t} | \mathcal{D}_t) \sim \mathcal{N} \mathcal{G}(m_{j,t}, C_{j,t}, n_{j,t}, s_{j,t})$
\begin{eqnarray*}
\theta_{j,t} | \lambda_{j,t}, \mathcal{D}_t  &\sim& \mathcal{N}\left( m_{j,t}, C_{j,t} / \left( s_{j,t-1} \lambda_{j,t} \right) \right) \, , \\
\lambda_{j,t} | \mathcal{D}_t &\sim& \mathcal{G}\left(n_{j,t}/2, n_{j,t} s_{j,t-1}/2 \right) \, .
\end{eqnarray*}
Hence with $y_{j,t}$ the observed data at $t$ the parameters follow the classic Kalman filter update:
\begin{eqnarray*}
e_{j,t} &=& y_{j,t} - F_{j,t}^T a_{j,t} \, , \\
q_{j,t} &=& s_{j,t-1} + F_{j,t}^T R_{j,t} F_{j,t} \, , \\
A_{j,t} &=& R_{j,t} F_{j,t} / q_{j,t} \, , \\
z_{j,t} &=& \left(r_{j,t} + e_{j,t}^2 / q_{j,t} \right) / \left( r_{j,t} + 1 \right) \, , \\
m_{j,t} &=& a_{j,t} + A_{j,t} e_{j,t} \, ,  \\
C_{j,t} &=& \left(R_{j,t} - A_{j,t} A_{j,t}^T q_{j,t}\right)z_{j,t} \, , \\
n_{j,t} &=& r_{j,t} + 1 \, ,\\
s_{j,t} &=& z_{j,t} s_{j,t-1} \, .
\end{eqnarray*}

\subsection{Decoupling with Variational Bayes}
The decoupling step approximate the multivariate posterior distribution by a product of Normal-Gamma distributions $p(\theta_{j,t},\lambda_{j,t} | \mathcal{D}_t) \sim \mathcal{N} \mathcal{G}(m_{j,t}, C_{j,t}, n_{j,t}, s_{j,t})$ whose parameters are obtain by minimising the Kullback-Leibler distance. 
\begin{eqnarray*}
m_{j,t} &=& E[\lambda_{j,t} \theta_{j,t}] / E[\lambda_{j,t}] \, , \\
V_{j,t} &=& E[\lambda_{j,t} \left( \theta_{j,t} - m_{j,t} \right) \left( \theta_{j,t} - m_{j,t} \right)^T] \, , \\
d_{j,t} &=& E[\lambda_{j,t} \left( \theta_{j,t} - m_{j,t} \right)^T V_{j,t}^{-1} \left( \theta_{j,t} - m_{j,t} \right)] \, , \\
s_{j,t} &=& \left(n_{j,t} + p_{j,t} - d_{j,t} \right) / \left( n_{j,t} E[\lambda_{j,t}] \right) \, , \\
C_{j,t} &=& s_{j,t} V_{j,t} \, .
\end{eqnarray*}
$n_{j,t}$ is updated through an optimisation step solving:
\begin{equation*}
log(n_{j,t} + p_{j,t} - d_{j,t}) - \psi(n_{j,t}/2) - (p_{j,t} - d_{j,t}) / n_{j,t} - log(2 E[\lambda_{j,t}]) + E[log(\lambda_{j,t})] = 0  \, .
\end{equation*}

\subsection{Evolution to $t+1$}
For the evolution from $t$ to $t+1$ the states $\theta_{j,t}$ follow a random walk $\theta_{j,t+1} = G_{j,t+1}\theta_{j,t} + \omega_{j,t}$. The coefficient matrix $G_{j,t}$ is updated with the parents down set values following $1 - \left( \Delta_T + 1 - l \right)^{-1}$ for $l=1:\Delta_T$. $m_{j,t}$ is updated to include a value for the new members of the up-set. With $B_{j,t+1} = G_{j,t+1}C_{j,t}G_{j,t+1}^T$ the covariance matrix follow the following block discounting update:
\begin{equation*}
W_{j,t+1} = 
\begin{pmatrix}
\left(\frac{1}{\delta_{j, \phi}} - 1 \right) B_{j,t+1}[:n_E, :n_E]  & \left(\frac{1}{\sqrt{\delta_{j, \phi} \delta{j, \gamma}}} - 1 \right) B_{j,t+1}[:n_E, n_E:] \\
0 &  \left(\frac{1}{\delta_{j, \gamma}} - 1 \right) B_{j,t+1}[n_E:, n_E:]
\end{pmatrix}  \, .
\end{equation*}
where $n_E$ corresponds to the number of external variables, vs. parents ones, $\delta_\phi$ the external variables factor and $\delta_\gamma$ the parental one. The filters' parameters follow:
\begin{eqnarray*}
a_{j,t+1} &=& G_{j,t+1} m_{j,t+1} \, , \\
R_{j,t+1} &=& B_{j,t+1} + W_{j,t+1} \, , \\
r_{j,t+1} &=& \beta_j n_{j,t} \, .
\end{eqnarray*}
where this last equation represent the beta-stochastic-volatility model. 

\subsection{Prior inference}
The prior for $t+1$ samples follow $p(\theta_{j,t+1},\lambda_{j,t+1} | \mathcal{D}_t) \sim \mathcal{N} \mathcal{G}(a_{j,t+1}, R_{j,t+1}, r_{j,t+1}, s_{j,t})$. Which are used to compute $\Gamma_{j,t+1} = [\theta_{j,t+1}]_{\forall j}$, $\Lambda_{j,t+1} = diag\left( \lambda_{0,t+1}, \dots,  \lambda_{N,t+1} \right)$ and $\mu_{t+1} = x_{j, t} \phi_{j,t+1}$. Recall $\theta_{j,t} = \left(\phi_{j,t}, \gamma_{j,t}\right)^T$. Hence we obtain the mean and covariance matrix to sample $y_{t+1}$:
\begin{eqnarray*}
A_{t+1} &=& \left(I - \Gamma_{t+1}\right)^{-1} \, , \\
\Sigma_{t+1} &=& \left( \left(I - \Gamma_{t+1}\right)^T \Lambda_{t+1}  \left(I - \Gamma_{t+1}\right) \right)^{-1} \, , \\
y_{t+1} &\sim& \mathcal{N}\left( A_{t+1} \mu_{t+1}, \Sigma_{t+1} \right) \, .
\end{eqnarray*}

%%%%%%%%%%%%%%%%%%%%%%%%%%%%%%%%%%%%%%%%%%%%%%%%%%%%%%%%%%%%%%%%%%%%%%%%%%%%%%%%%%%%
\section{BibTeX}
\bibliographystyle{jabes}

\bibliography{bibTexLib_HSGDLM}

\begin{thebibliography}{27}
\newcommand{\enquote}[1]{``#1''}
\expandafter\ifx\csname natexlab\endcsname\relax\def\natexlab#1{#1}\fi
\expandafter\ifx\csname url\endcsname\relax
  \def\url#1{\texttt{#1}}\fi
\expandafter\ifx\csname urlprefix\endcsname\relax\def\urlprefix{URL }\fi

\bibitem[{Anderson et~al.(2000)Anderson, Bollerslev, Diebold, and
  Labys}]{Anonymous:neN5kQNv}
Anderson, T., Bollerslev, T., Diebold, F., and Labys, P. (2000),
  \enquote{{Great realisations},} \textit{Risk},
  \urlprefix\url{http://scholar.google.comjavascript:void(0)}.

\bibitem[{Borland(1998)}]{Borland:1998wf}
Borland, L. (1998), \enquote{{Microscopic dynamics of the nonlinear
  Fokker-Planck equation: A phenomenological model },} \textit{physical
  review}, 57.

\bibitem[{Borland(2002)}]{Borland:2010km}
--- (2002), \enquote{{A theory of non-Gaussian option pricing},}
  \textit{Quantitative Finance},
  \urlprefix\url{http://www.tandfonline.com/doi/abs/10.1080/14697688.2002.0000009}.

\bibitem[{Borland(2004)}]{Borland:2004wu}
--- (2004), \enquote{{A multi-time scale non-Gaussian model of stock returns},}
  \textit{arXiv.org}, \urlprefix\url{http://arxiv.org/abs/cond-mat/0412526v3}.

\bibitem[{Borland and Bouchaud(2005)}]{Borland:2005tk}
Borland, L. and Bouchaud, J.-P. (2005), \enquote{{On a multi-timescale
  statistical feedback model for volatility fluctuations},} \textit{arXiv.org},
  \urlprefix\url{http://arxiv.org/abs/physics/0507073v1}.

\bibitem[{Borland et~al.(2005)Borland, Bouchaud, Muzy, and
  Zumbach}]{Borland:2005vd}
Borland, L., Bouchaud, J.-P., Muzy, J.-F., and Zumbach, G. (2005),
  \enquote{{The Dynamics of Financial Markets--Mandelbrot's multifractal
  cascades, and beyond},} \textit{arXiv.org},
  \urlprefix\url{https://arxiv.org/abs/cond-mat/0501292}.

\bibitem[{Calvet and Fisher(2001)}]{Calvet:2001fo}
Calvet, L.~E. and Fisher, A.~J. (2001), \enquote{{Forecasting multifractal
  volatility},} \textit{Journal of Econometrics}, 105, 27--58,
  \urlprefix\url{http://linkinghub.elsevier.com/retrieve/pii/S0304407601000690}.

\bibitem[{Calvet and Fisher(2004)}]{Calvet:2004fm}
--- (2004), \enquote{{How to Forecast Long-Run Volatility: Regime Switching and
  the Estimation of Multifractal Processes},} \textit{Journal of Financial
  Econometrics}, 2, 49--83,
  \urlprefix\url{http://jfec.oxfordjournals.org/content/2/1/49.short}.

\bibitem[{Corsi(2004)}]{Corsi:2004eg}
Corsi, F. (2004), \enquote{{A Simple Long Memory Model of Realized
  Volatility},} \textit{SSRN Electronic Journal},
  \urlprefix\url{http://www.ssrn.com/abstract=626064}.

\bibitem[{Corsi et~al.(2008)Corsi, Mittnik, Pigorsch, and
  Pigorsch}]{Corsi:2008ei}
Corsi, F., Mittnik, S., Pigorsch, C., and Pigorsch, U. (2008), \enquote{{The
  Volatility of Realized Volatility},} \textit{Econometric Reviews}, 27,
  46--78,
  \urlprefix\url{http://www.tandfonline.com/doi/abs/10.1080/07474930701853616}.

\bibitem[{Corsi and Reno(2009)}]{Corsi:2009ue}
Corsi, F. and Reno, R. (2009), \enquote{{HAR volatility modelling with
  heterogeneous leverage and jumps},} \textit{Available at SSRN 1316953},
  \urlprefix\url{http://citeseerx.ist.psu.edu/viewdoc/download?doi=10.1.1.435.4217&rep=rep1&type=pdf}.

\bibitem[{Dacorogna et~al.(1997)Dacorogna, M{\"u}ller, Pictet, and
  Olsen}]{Dacorogna:1997ez}
Dacorogna, M.~M., M{\"u}ller, U.~A., Pictet, O.~V., and Olsen, R.~B. (1997),
  \enquote{{Modelling Short-Term Volatility with GARCH and HARCH Models},}
  \textit{SSRN Electronic Journal},
  \urlprefix\url{http://www.ssrn.com/abstract=36960}.

\bibitem[{Diebold et~al.(1997)Diebold, Hickman, Inoue, and
  Schuermann}]{fxd:1997ul}
Diebold, F.~X., Hickman, A., Inoue, A., and Schuermann, T. (1997),
  \textit{{Converting 1-day volatility to h-day volatility: Scaling by sqrt(n)
  is worse than you think}}, F. Diebold,
  \urlprefix\url{http://scholar.google.com/scholar?q=related:GrOEHFKqSLQJ:scholar.google.com/&hl=en&num=20&as_sdt=0,5}.

\bibitem[{Gruber and West(2016{\natexlab{a}})}]{Gruber:2016wn}
Gruber, L.~F. and West, M. (2016{\natexlab{a}}), \enquote{{Bayesian forecasting
  and scalable multivariate volatility analysis using simultaneous graphical
  dynamic models},} \textit{arXiv.org},
  \urlprefix\url{http://arxiv.org/abs/1606.08291v1}.

\bibitem[{Gruber and West(2016{\natexlab{b}})}]{Gruber:2016il}
--- (2016{\natexlab{b}}), \enquote{{GPU-Accelerated Bayesian Learning and
  Forecasting in Simultaneous Graphical Dynamic Linear Models},}
  \textit{Bayesian Analysis}, 11, 125--149,
  \urlprefix\url{http://projecteuclid.org/euclid.ba/1425304898}.

\bibitem[{Hull(2017)}]{Hull:2019wk}
Hull, J.~C. (2017), \enquote{{The Greek Letters},} in \textit{Options, Futures,
  and Other Derivatives}, ed. P.~E. Limited, pp. 421--452,
  \urlprefix\url{https://ebookcentral.proquest.com/lib/imperial/detail.action?docID=5186416}.

\bibitem[{Johannes et~al.(2005)Johannes, Polson, and Stroud}]{mj:2006vf}
Johannes, M., Polson, N., and Stroud, J. (2005), \enquote{{Sequential parameter
  estimation in stochastic volatility models with jumps},}
  \textit{columbia.edu},
  \urlprefix\url{http://scholar.google.com/scholar?q=related:fCswAqOxaHAJ:scholar.google.com/&hl=en&num=20&as_sdt=0,5}.

\bibitem[{LeBaron(2001)}]{LeBaron:2001cp}
LeBaron, B. (2001), \enquote{{Stochastic volatility as a simple generator of
  apparent financial power laws and long memory},} \textit{Quantitative
  Finance}, 1, 621--631,
  \urlprefix\url{http://www.tandfonline.com/doi/abs/10.1088/1469-7688/1/6/304}.

\bibitem[{Lynch and Zumbach(2003)}]{Lynch:2003cn}
Lynch, P.~E. and Zumbach, G. (2003), \enquote{{Market heterogeneities and the
  causal structure of volatility},} \textit{Quantitative Finance}, 3, 320--331,
  \urlprefix\url{http://www.tandfonline.com/doi/abs/10.1088/1469-7688/3/4/308}.

\bibitem[{McAlinn and West(2016)}]{McAlinn:2016tm}
McAlinn, K. and West, M. (2016), \enquote{{Dynamic Bayesian Predictive
  Synthesis in Time Series Forecasting},} \textit{arXiv.org},
  \urlprefix\url{http://arxiv.org/abs/1601.07463v3}.

\bibitem[{M{\"u}ller et~al.(1997)M{\"u}ller, Dacorogna, Dav{\'e}, Olsen,
  Pictet, and von Weizs{\"a}cker}]{Muller:1997he}
M{\"u}ller, U.~A., Dacorogna, M.~M., Dav{\'e}, R.~D., Olsen, R.~B., Pictet,
  O.~V., and von Weizs{\"a}cker, J.~E. (1997), \enquote{{Volatilities of
  different time resolutions {\textemdash} Analyzing the dynamics of market
  components},} \textit{Journal of Empirical Finance}, 4, 213--239,
  \urlprefix\url{http://linkinghub.elsevier.com/retrieve/pii/S0927539897000078}.

\bibitem[{Muzy et~al.(2000)Muzy, Delour, and Bacry}]{Muzy:2000cz}
Muzy, J.-F., Delour, J., and Bacry, E. (2000), \enquote{{Modelling fluctuations
  of financial time series: from cascade process to stochastic volatility
  model},} \textit{arXiv.org}, 537--548,
  \urlprefix\url{http://link.springer.com/10.1007/s100510070131}.

\bibitem[{Nakajima(2014)}]{Nakajima:2014dt}
Nakajima, J. (2014), \enquote{{Bayesian Analysis of Multivariate Stochastic
  Volatility with Skew Return Distribution},} \textit{Econometric Reviews}, 8,
  1--23,
  \urlprefix\url{http://www.tandfonline.com/doi/full/10.1080/07474938.2014.977093}.

\bibitem[{Noureldin et~al.(2011)Noureldin, Shephard, and
  Sheppard}]{Noureldin:2011bc}
Noureldin, D., Shephard, N., and Sheppard, K. (2011), \enquote{{Multivariate
  high-frequency-based volatility (HEAVY) models},} \textit{Journal of Applied
  Econometrics}, 27, 907--933,
  \urlprefix\url{http://doi.wiley.com/10.1002/jae.1260}.

\bibitem[{Prado and West(2010)}]{Prado:2010wu}
Prado, R. and West, M. (2010), \textit{{Time series: modeling, computation, and
  inference}}, Chapman {\&} Hall Book CRC Press,
  \urlprefix\url{https://books.google.com/books?hl=en&lr=&id=kUhKLLdGGZ4C&oi=fnd&pg=PP1&dq=Time+series+modeling+computation+and+inference&ots=F0XERR9f6n&sig=0O4heSO6OQIEAi_9PbxqhRAlRlI}.

\bibitem[{Zhao et~al.(2016)Zhao, Xie, and West}]{Zhao:2016kr}
Zhao, Z.~Y., Xie, M., and West, M. (2016), \enquote{{Dynamic dependence
  networks: Financial time series forecasting and portfolio decisions},}
  \textit{Applied Stochastic Models in Business and Industry}, 32, 311--332,
  \urlprefix\url{http://doi.wiley.com/10.1002/asmb.2161}.

\bibitem[{Zumbach and Lynch(2001)}]{Zumbach:2001iz}
Zumbach, G. and Lynch, P. (2001), \enquote{{Heterogeneous volatility cascade in
  financial markets},} \textit{Physica A: Statistical Mechanics and its
  Applications}, 298, 521--529,
  \urlprefix\url{http://linkinghub.elsevier.com/retrieve/pii/S0378437101002497}.

\end{thebibliography}

%\listoffigures
\end{document}